\documentclass[12pt]{article}
\usepackage{a4,epsfig}
\usepackage{amsmath}
\usepackage{graphics}
\newcounter{condition}[subsection] 
\renewcommand{\thecondition}{\arabic{section}.\arabic{subsection}.%
                             \arabic{condition}}
\newcommand{\BeginCon}{\vspace{0.5\baselineskip}}
\newcommand{\Condition}{\vspace{0.5\baselineskip}%
      \refstepcounter{condition}%
      \noindent{\bfseries Condition \thecondition.}\hspace{10pt}}
\newcommand{\EndCon}{\vspace{1.0\baselineskip}}			
\newcommand{\Con}[1]{Condition~\ref{#1}}
\newcommand{\beeq}{\begin{equation}}
\newcommand{\eneq}{\end{equation}}
\newcommand{\beeqa}{\begin{eqnarray}}
\newcommand{\eneqa}{\end{eqnarray}}
\newcommand{\nn}{\nonumber}
\newcommand{\nl}{\nonumber \\}
\newcommand{\suml}{\sum\limits}

\newcommand{\eqn}[1]{Eq.(\ref{#1})}
\newcommand{\fig}[1]{Fig.\ref{#1}}
\newcommand{\Sec}[1]{Section~\ref{#1}}
\newcommand{\vhi}{\varphi}
\newcommand{\al}{\alpha}
\newcommand{\be}{\beta}
\newcommand{\la}{\lambda}

\newcommand{\plaat}[3]%
          {\raisebox{#3pt}{\epsfig{figure=epsfiles/#1.eps,width=#2pt}}}

\newcommand{\ddv}[1]{\frac{\partial}{\partial #1}}
\newcommand{\ddmu}{\ddv{\mm}}
\newcommand{\Ai}{\mbox{Ai}}
\newcommand{\Bi}{\mbox{Bi}}
\newcommand{\SD}{\mbox{SD}}
\newcommand{\Stw}{\mbox{S2}}
\newcommand{\Sth}{\mbox{S3}}
\newcommand{\Sfo}{\mbox{S4}}
\newcommand{\sfrac}[2]{{\textstyle\frac{#1}{#2}}}
\newcommand{\C}[1]{C_{#1}}
\newcommand{\mm}{\mu}
\newcommand{\lat}{\lambda_3}
\newcommand{\laf}{\lambda_4}
\newcommand{\gt}{g_3}
\newcommand{\gf}{g_4}

\newcommand{\half}{{\frac{1}{2}}}
\newcommand{\sixt}{{\frac{1}{6}}}
\newcommand{\twft}{{\frac{1}{24}}}
\newcommand{\vhitf}{\vhi^3\!+\!\vhi^4}
\newcommand{\vhib}{\bar{\vhi}}
\newcommand{\xx}{x}
\newcommand{\xb}{\bar{\xx}}
\newcommand{\df}{:=}
\newcommand{\pa}{\partial}
\newcommand{\pb}{\bar{\partial}}
\newcommand{\phb}{\bar{\phi}}

\begin{document}

\thispagestyle{empty}
\vspace*{2cm}
\begin{center}
%
%
%
{\bf\Huge Zero-dimensional field theory}

\vspace{2\baselineskip}
%
%

{\large 
E.N.~Argyres$\,^a$, 
A.F.W.~van Hameren$\,^{a\,b\,\ddagger}$,
R.H.P.~Kleiss$^{\,b\,\star}$, 
\\[.2cm] 
and 
C.G.~Papadopoulos$^{\,a\,+}$}
\\[1cm]
$^a$ Institute of Nuclear Physics, NCSR ``Demokritos'', 15310, Athens, Greece
\\
$^b$ University of Nijmegen, Nijmegen, The Netherlands
\vspace{0.25\baselineskip}
%
%

\vspace{3\baselineskip}
%
%
{\bf Abstract}

\vspace{\baselineskip}
\parbox{0.85\linewidth}{\hspace*{15pt}%
%
A study
of zero-dimensional theories, based on exact results, is presented.
First, relying on a simple diagrammatic representation
of the theory, equations involving the generating function 
of all connected Green's
functions are constructed. Second, exact  solutions of these equations 
are obtained for several theories. 
Finally, renormalization is carried out. Based on the anticipated
knowledge of the exact solutions 
the full dependence on the renormalized coupling constant
is studied.
}
\end{center}
\vspace*{\fill}
\rule{7cm}{1pt}\\
$^\ddagger$ {\tt andrevh@sci.kun.nl},
$^\star$ {\tt kleiss@sci.kun.nl}, 
$^+$ {\tt Costas.Papadopoulos@cern.ch}. 
\newpage
%
%
\section{Introduction}
In this paper we study several aspects of zero-dimensional quantum field
theory. Such theories may serve as a model (the static ultra-local limit) of
more realistic quantum field theories, and as a useful didactic object in their
own right, since zero-dimensional theories, for which the path integral is
actually a simple integral, allow for many explicit and exact solutions that
cannot be obtained in higher dimensions. As recent examples, we may quote 't
Hooft \cite{hooft} and Bender {\it et al.\/}~\cite{bender}. Questions of
particular interest here are the behavior of theories in high orders of
perturbation theory (either many loops, or large number of external legs), and
of the relation between the diagrammatic perturbation expansion and the full
solution. The layout of this paper is as follows. We start by a diagrammatic
(re)derivation of equations that govern the set of all connected Green's
functions of the theory. We show how for a general scalar theory with arbitrary
interactions the Green's functions may be obtained order by order. We point out
how the Schwinger-Dyson equation, although derivable from purely diagrammatic
arguments, in fact describes a much larger class of solutions. Next, we discuss
the representation of these solutions as path integrals over contours in the
complex $\vhi$-plane. Exact solutions for theories with interactions up to
$\vhi^4$ are obtained including explicitly perturbative and non-perturbative
contributions. We show how a classification of the allowed contours in the
complex $\vhi$-plane can immediately determine properties of the
non-perturbative character of these theories. Renormalization, which for these
theories is equivalent to imposing restrictions to diagrams, is also studied.
The wave function renormalizations are fully determined and their dependence on
the renormalized coupling constant of the theory is presented and discussed. 
 
\section{Basic equations}

In this section we derive equations for an arbitrary zero-dimensional field
theory. The derivation is based entirely upon the diagrammatic representation
of the theory. A theory is diagrammatically defined by a sequence of vertices
that are weighted by the `coupling' constants taken for convenience as
$-\lambda_k$, for the $k$-th vertex. In fact, the two-point coupling
$\lambda_2=m^2\equiv \mm$ can be eliminated by the introduction of the
propagator which means that every line of a diagram accounts for a factor
$1/\mm$. Moreover a loop in a graph is counted by an additional parameter,
$\hbar$. A solution of a zero-dimensional theory is determined by a sequence
of objects $\C{n}$, $n=0,1,2,\ldots$ that represent the connected, $n$-point
Green's functions, i.e. the sum of all connected diagrams with $n$ external
lines. One can define the generating function of the Green's functions as
\begin{equation}
   \phi(x) = \suml_{n=0}^\infty\frac{x^n}{n!}\C{n+1} \;\;.
\label{phi}
\end{equation}
We want to write down an equation for $\phi$, and in order to do so, we
represent it with a diagram:
\begin{equation}
  \phi(x)  = \plaat{f0}{33}{-5.5}\;\;.
\nn\end{equation}
Derivatives of $\phi$ with respect to $x$ are represented by extra lines, 
\begin{equation}
  \phi'(x) = \plaat{f1}{50}{-6}\;\;\;,\;\;\;
  \phi''(x) = \plaat{f2}{46}{-9}\;\;,
\nn\end{equation}
and so on.

\subsection{The Schwinger-Dyson equation}

Let us consider a theory with a $k$-th vertex. In order to write
an equation for $\phi$ we start with the bare vertex and we attach $k-1$ blobs 
\begin{equation}
  \plaat{kzero}{45}{-24} \quad\quad \hbar^0 \frac{\phi^{k-1}}{(k-1)!}\;.
  \nn
\end{equation}
The factor $1/(k-1)!$ is due to the $k-1$ identical blobs.
Considering a one loop attachment we similarly have 
\begin{equation}
  \plaat{kone}{45}{-24} \quad\quad \hbar^1 \frac{\phi'}{2!} 
  \frac{\phi^{k-3}}{(k-3)!}\;.
  \nn
\end{equation}
The factor $1/(k-2)!$ is again due to the $k-2$ identical blobs, whereas the
$1/2!$ is due to the symmetry factor of blob with two lines. Following this
reasoning we can proceed with higher powers of $\hbar$. For instance at the
two-loop level we get two terms
\begin{alignat}{2}
  &\plaat{ktwoa}{45}{-24} 
  &\quad\quad& \hbar^2 \frac{\phi^{'2}}{2!2!2!} 
   \frac{\phi^{k-5}}{(k-5)!}
   \nn\\
  &\plaat{ktwob}{45}{-24}
  &\quad\quad& \hbar^2 \frac{\phi''}{3!} 
   \frac{\phi^{k-4}}{(k-4)!}\;.
   \nn
\end{alignat}
Finally the last term, i.e the term with the largest number of loop
attachments, will simply read 
\begin{equation}
  \plaat{klast}{42}{-9}\quad\quad \hbar^{k-2} \frac{\phi^{(k-2)}}{(k-1)!} \;.
  \nn
\end{equation}
The result looks like
\begin{equation}
\plaat{sd1}{33}{-6}= \plaat{sd2}{20}{1.5}
  + \plaat{kzero}{45}{-24}
  +  \plaat{kone}{45}{-24}
  + \plaat{ktwoa}{45}{-24}
  + \plaat{ktwob}{45}{-24}
  + \ldots
  + \plaat{klast}{42}{-9} \;\;.
\nn\end{equation}
The equation reads
\begin{multline}
  x=\mm\phi
  +\lambda_k\bigg(  
    \frac{\phi^{k-1}}{(k-1)!}
  + \hbar^1\frac{\phi'}{2!}\frac{\phi^{k-3}}{(k-3)!}\\
  + \hbar^2\frac{\phi^{'2}}{2!2!2!}\frac{\phi^{k-5}}{(k-5)!}
  + \hbar^2\frac{\phi''}{3!}\frac{\phi^{k-4}}{(k-4)!}
  + \ldots
  + \hbar^{k-2} \frac{\phi^{(k-2)}}{(k-1)!}\bigg) \;.
  \label{gsd}
\end{multline}
For an arbitrary theory a sum over $k$ should be understood. It represents a
non-linear differential equation for $\phi$, the Schwinger-Dyson (\SD)
equation,  which has been derived by the direct application of the Feynman
rules.

In order to be more specific let us consider a theory with only a $3$-point and
a $4$-point vertex. Following the abovementioned reasoning a diagrammatic
equation for $\phi$ looks like
\begin{equation}
  \plaat{f0}{33}{-5.5} \;=\; x\plaat{sd2}{20}{1.5} 
                       \;+\; \plaat{sd3}{40}{-21} 
                       \;+\; \plaat{sd4}{42}{-5.5}  
                       \;+\; \plaat{sd5}{50}{-22} 
                       \;+\; \plaat{sd6}{41}{-20} 
                       \;+\; \plaat{sd7}{52}{-6}\;\;.
\nn\end{equation}
This reads
\begin{equation}
  \phi(x) =   \frac{x}{\mm} 
            - \frac{\lat}{2\mm}\left[\phi(x)^2 + \hbar\phi'(x)\right] 
	    - \frac{\laf}{6\mm}\left[\phi(x)^3 + 3\hbar\phi(x)\phi'(x) 
	                                      + \hbar^2\phi''(x)\right]\;\;.
\label{sd}\end{equation}
This equation generates equations for the Green's functions if the power series
of $\phi(x)$ is inserted. The first three are
\begin{align}
   \C{1} =& - \frac{1}{6\mm}\C{1}^2(\laf\C{1}+3\lat)
          - \frac{\hbar}{2\mm}\C{2}(\laf\C{1}+\lat)
          - \frac{\hbar^2}{6\mm}\laf\C{3}\;\;,\nl
   \C{2} =& - \frac{1}{2\mm}(2\lat\C{1}\C{2} - 2 + \laf\C{1}^2\C{2})
          - \frac{\hbar}{2\mm}(\lat\C{3} + \laf\C{1}\C{3} + \laf\C{2}^2) 
          - \frac{\hbar^2}{6\mm}\laf\C{4}\;\;,\nl
   \C{3} =& - \frac{1}{2\mm}(2\lat\C{1}\C{3} + 2\lat\C{2}^2 + \laf\C{3}\C{1}^2 
          + 2\laf\C{2}^2\C{1})\nl
          &- \frac{\hbar}{2\mm}(3\laf\C{2}\C{3} + \lat\C{4} + \laf\C{1}\C{4})
          - \frac{\hbar^2}{6\mm}\laf\C{5}\;\;.
\label{cfromsd}
\end{align}

The \SD\ equation is invariant under certain redefinition of the parameters
involved. It is not difficult to prove that if $\phi(\mm,\lambda_k,\hbar;x)$ is
a solution, also
$c^{\be}\phi(c^{\al-2\be}\mm,c^{\al-k\be}\lambda_k,c^{\al}\hbar; c^{\al-\be}x)$
is a solution for any $c,\al,\be$. This scaling property is also a concequence
of the fact that $\phi(\mm,\lambda_k,\hbar;x)/\sqrt{\hbar/\mm}$ is a
dimensionless function of the scaled variables $y=x/\sqrt{\hbar\mm}$ and
$g_k=\lambda_k\hbar^{k/2-1}/\mm^{k/2}$. The scaling property can be expressed
with the following equations, derived, for instance, by differentiating with
respect to $c$ and taking $c=1$:
\begin{align}
  \left(  x\ddv{x} + \mm\ddv{\mm} + \hbar\ddv{\hbar}
        + \lambda_k\ddv{\lambda_k} \right)\phi &= 0\;\;,\nl
  \left(  1 + x\ddv{x} + 2\mm\ddv{\mm} + k\lambda_k\ddv{\lambda_k} 
        \right)\phi &=  0\;\;.
\label{counting}
\end{align}
These equations are equivalent to the usual topological relations that relate
the number of external lines $E$, the number of internal lines $I$, the number
of $k$-vertices $V_k$, and the number of loops $L$, appearing in any diagram,
\[
k V_k=E+2I\quad\quad V_k=I+1-L\;.
\]
A sum over $k$ should be undesrtood in the general case. 

\subsection{Stepping equations}

In the diagrammatic construction, 
one assumes that every Green's function can be
written as a sum of diagrams, consisting of vertices connected by lines
(propagators). 
The power of $1/\mm$ in a diagram is equal to the number of propagators,
and hence the operation $-\partial/\partial\mm$ on this diagram corresponds to
cutting a single propagator in all possible places in that diagram. There are
two possibilities for the result: the chosen propagator may be either 
a part of a loop,
in which case the diagram remains connected
when we cut this line, or part of the `tree skeleton', such that cutting it
makes the diagram diconnected:
\begin{equation}
  \plaat{f0}{33}{-6} \;=\; \plaat{s21}{40}{-6} \;+\; \plaat{s22}{58}{-6}\;\;.
\nn\end{equation}
In the first case the cut diagram remains connected but gains two external
lines at the price of one loop ({\it i.e.\/} on power of $\hbar$); in the
second place, the cut diagram falls apart into two connected diagrams:
\begin{equation}
  \ddmu\plaat{f0}{33}{-6} \;=\; \plaat{s21c}{38}{-6} 
                          \;+\; \plaat{s22c}{68}{-6}\;\;.
\nn\end{equation}
Putting in the correct symmetry factors, we can express this procedure by
\begin{equation}
\mbox{S2:\hspace*{2cm}} 
\frac{\partial}{\partial\mm}\phi(x) + \phi(x)\phi'(x) 
                                      + \frac{\hbar}{2}\phi''(x) = 0\;\;,
\label{s2}
\end{equation}
where the second term comes from diagrams that fall apart under cutting, and
the third one from loops that are cut open. We call \eqn{s2} the Step-2
equation (\Stw) since it describes a procedure in which the number of external
legs is increased in steps of 2.

Like \SD\ (\eqn{sd}), the Stepping equation S2 implies relations
between various $\C{}$'s. The lowest few of these read
\begin{align}
  \C{3} &= -\frac{2}{\hbar}\left(\C{1}\C{2} + \ddmu\C{1}\right)\;\;,\nl
  \C{4} &= -\frac{2}{\hbar}\left(\C{1}\C{3} + \C{2}^2 
                                             + \ddmu\C{2}\right)\;\;,\nl
  \C{5} &= -\frac{2}{\hbar}\left(\C{1}\C{4} + 3\C{2}\C{3} 
                                             + \ddmu\C{3}\right)\;\;,\nl
  \C{6} &= -\frac{2}{\hbar}\left(\C{1}\C{5} + 4\C{2}\C{4} + \C{3}^2 
                                             + \ddmu\C{4}\right)\;\;,
\label{cfroms2}\end{align}
and so on. Note that S2 is completely independent of the interaction
potential, and therefore perforce contains information independent of that
contained in the \SD. It follows that there must be solutions to \SD\ that do
{\em not\/} obey \Stw\, and that these solutions cannot be represented by
Feynman diagrams.

It is possible to combine \SD\ and \Stw\, in the following manner. Taking the
first equation in \eqn{sd}, we express $\C{3}$ in $\C{2}$ and $\C{1}$, and
solve for $\C{2}$:
\begin{equation}
  \C{2} = \frac{  2\laf{\displaystyle\frac{\partial\C{1}}{\partial\mm}
                - \frac{1}{\hbar}\left(6\mm \C{1} 
		+ 3\lat\C{1}^2 + \laf\C{1}^3\right)}}
	       {3\lat + \laf\C{1}}\;\;.
\nn\end{equation}
Inserting this into the second equation of \eqn{sd}, we find
a differential equation for $\C{1}$ alone:
\begin{align}
  0 &= 4\hbar^2\laf^3\left(\frac{\partial\C{1}}{\partial\mm}\right)^2 
        -2\hbar^2\laf^2\frac{\partial^2\C{1}}{\partial\mm^2}
        \left(\laf\C{1} + 3\lat\right)\nl
    &- \hbar\frac{\partial\C{1}}{\partial\mm}
        \left[9\lat^2(3\lat + \laf\C{1}) + 6\mm \laf^2\C{1} 
	                             - 36\mm \lat\laf\right]\nl
    &- 9\mm \lat\C{1}^2(3\lat + \laf\C{1}) + 3\hbar \laf^2\C{1}^2
        - 54\mm^2\lat\C{1} - 27\hbar \lat^2 \;\;.
\label{c1}\end{align}
By inserting the series expansion
\begin{equation}
  \C{1} = \suml_{k\ge1} \alpha_k\hbar^k\;\;,
\nn\end{equation}
we can then successively determine the coefficients:
\begin{align}
  \alpha_1 &= -\frac{1}{2\mm^2}\lat\;\;,\nl
  \alpha_2 &= -\frac{1}{24\mm^5}(15\lat^3 - 16\mm \lat\laf)\;\;,\nl
  \alpha_3 &= -\frac{1}{48\mm^8}(90\lat^5 - 185\mm \lat^3\laf 
                                          + 66\mm^2\lat\laf^2)\;\;,\nl
  \alpha_4 &= -\frac{1}{1152\mm^{11}}(9945\lat^7 - 30270\mm \lat^5\laf 
                                                 + 24280\mm^2\lat^3\laf^2 
						 - 4352\mm^3\lat\laf^3)\;\;,
\nonumber
\end{align}
and so on.

Whereas \Stw\ is independent of the interaction potential, we can also derive
stepping equations by deleting vertices rather than cutting lines.
For example, let us depict all possible ways in which a selected $\vhi^3$
vertex (denoted by a dot) can occur in a connected graph:
\begin{equation}
  \plaat{f0}{33}{-6} \;=\; \plaat{s31}{50}{-7} 
                     \;+\; \plaat{s32}{70}{-6.5} 
                     \;+\; \plaat{s33}{70}{-7} 
		     \;+\; \plaat{s34}{66}{-23.5}\;\;.
\nn\end{equation}
Deleting this vertex gives us
\begin{equation}
  -\ddv{\lat}\plaat{f0}{33}{-6} \;=\; \plaat{s31c}{40}{-9.5} 
                               \;+\; \plaat{s32c}{70}{-6.5} 
                               \;+\; \plaat{s33c}{70}{-9} 
			       \;+\; \plaat{s34c}{70}{-19.5}\;\;.
\nn\end{equation}
or, in terms of $\phi(x)$, the following Step-3 equation (\Sth):
\begin{equation}
\mbox{S3:\hspace*{2cm}}  
\frac{\partial}{\partial\lat}\phi + \frac{1}{6}\hbar^2\phi''' 
  + \frac{1}{2}\hbar\left[\phi\phi'' + (\phi')^2\right] 
  + \frac{1}{2}\phi^2\phi' = 0\;\;.
\label{s3}
\end{equation}
A similar treatment holds for $\vhi^4$ vertices: the possible ways in which
such a vertex can occur is given by
\beeqa
  \plaat{f0}{33}{-6} \;=\; \plaat{s41}{40}{-7.5} 
                      &+&  \plaat{s42}{65}{-6} 
                     \;+\; \plaat{s43}{65}{-6}
                     \;+\; \plaat{s44}{62}{-15}\nl 
		      &+&  \plaat{s45}{65}{-6}
                     \;+\; \plaat{s46}{60}{-21} 
		     \;+\; \plaat{s47}{65}{-21.5} \;\;,
\nn\eneqa
and the result of
deleting is given by
\beeqa
-\ddv{\laf}\plaat{f0}{33}{-6} \;=\;
        \plaat{s41c}{38}{-11} &+& \plaat{s42c}{70}{-7.5} 
	                       \;+\; \plaat{s43c}{60}{-6}
                               \;+\; \plaat{s44c}{56}{-12.5}\nl 
			        &+& \plaat{s45c}{65}{-8.5} 
                               \;+\; \plaat{s46c}{60}{-20} 
			       \;+\; \plaat{s47c}{65}{-21.5} \;\;.
\nn\eneqa
The corresponding Step-4 equation (\Sfo) is
\begin{equation}
\mbox{S4:\hspace*{0.5cm}}
  \frac{\partial}{\partial \laf}\phi 
  + \frac{1}{24}\hbar^3\phi'''' 
  + \frac{1}{12}\hbar^2\left[2\phi\phi''' + 5\phi'\phi''\right] 
  + \frac{1}{4}\hbar\left[\phi^2\phi'' + 2\phi(\phi')^2\right] 
  + \frac{1}{6}\phi^3\phi' = 0\;\;.
\label{s4}
\end{equation}


\subsection{The charged scalar field}

Up to now we dealt with diagrammatic construction of zero-dimensional 
field theories involving only one field. As an illustrative extension, 
we consider a theory 
with two fields, {\it i.e.} a complex, or charged, scalar field. 
The Green's
functions are labeled with two integers, and the generating function has two
expansion parameters $\xx$ and $\xb$. 
Let us introduce the notation 
\begin{equation}
   \pa\df\frac{\pa}{\pa\xx} \qquad,\qquad \pb\df\frac{\pa}{\pa\xb} \;\;,
\nn\end{equation}
then
\begin{equation}
   \phi(\xx,\xb) = \sum_{n,m=0}^\infty\frac{\xx^n}{n!}\frac{\xb^m}{m!}\,
                                         \C{n,m+1}
\;\;.		
\nn\end{equation}
To write down the \SD\ equation, we introduce two kind of lines, 
distinguishable
by an arrow. The generating function is represented by 
\begin{equation}
   \phi(\xx,\xb) = \plaat{ch1}{30}{-5.5}       \qquad,\qquad
   \bar{\phi}(\xx,\xb) = \plaat{ch9}{30}{-5.5}\;\;.
\nn\end{equation}
An incoming external line represents a $\pb$, and an outgoing line represents
a $\pa$. Notice that 
\begin{equation}
   \hbar\pb\phi = \hbar\pa\phb = \plaat{ch7}{43}{-5} \;\;.
\nn\end{equation}
We also introduce a four point vertex with two 
incoming and two outgoing lines, 
so that the \SD\ equation we want $\phi$ to satisfy is given by 
\begin{equation}
  \plaat{ch1}{30}{-5.5} \;=\; x\plaat{ch2}{24}{-1} 
                       \;+\; \plaat{ch3}{45}{-24} 
                       \;+\; \plaat{ch4}{41}{-22.0}  
                       \;+\; \plaat{ch5}{41}{-22} 
                       \;+\; \plaat{ch6}{49}{-9}\;\;, 
\nn\end{equation}
or
\begin{equation}
   \phi = \frac{x}{\mu} - \frac{\la}{2\mu}\,\phi^2\phb 
                        - \frac{\la\hbar}{\mu}\,\phi\pa\phi
                        - \frac{\la\hbar}{2\mu}\,\phb\pb\phi
			- \frac{\la\hbar^2}{2\mu}\,\pa\pb\phi \;\;.
\end{equation}
Notice that incoming and outgoing lines are not equivalent, which is
represented by the symmetry factors. 

Also for the charged scalar field, we can write down stepping equations. For 
the first one, we use that in the diagrammatic interpretation
\begin{equation}
    \plaat{ch1}{30}{-5.5} =   \plaat{ch10}{40}{-5}  
                            + \plaat{ch11}{60}{-5}
                            + \plaat{ch12}{60}{-5}  \;\;,
\nn\end{equation}
leading to 
\begin{equation}
   \frac{\pa}{\pa\mu}\,\phi 
   = -\hbar\pa\pb\phi - \phi\pa\phi - \phb\pb\phi \;\;.
\end{equation}
The diagrammatic derivation of the stepping equation involving 
the derivative 
with respect to $\lambda$, although equally straightforward,
is rather cumbersome, leading to many terms which we refrain from listing here.

\section{Solutions to the equations}

\subsection{The integral representation}

The \SD\ equation is  highly non-linear. Let us consider the general term
connected with the coupling $\lambda_k$ in \eqn{gsd}:
\begin{equation}
Q_k= \frac{\phi^{k-1}}{(k-1)!}
+\hbar^1 \frac{\phi'}{2!} 
\frac{\phi^{k-3}}{(k-3)!}
+
\hbar^2 \frac{\phi^{'2}}{2!2!2!} 
\frac{\phi^{k-5}}{(k-5)!}
+
\hbar^2 \frac{\phi''}{3!} 
\frac{\phi^{k-4}}{(k-4)!}
+
\ldots
+
\hbar^{k-2} \frac{\phi^{(k-2)}}{(k-1)!} \;\;.
\nn\end{equation}
It can obviously be organized such that it can be written as 
\begin{equation}
Q_k= 
\sum_{m=1}^{k-1}\sum_{\{\vec{a}_{k-1};m\}} \frac{\hbar^{k-m-1}}{
(1!)^{a_1}a_1!(2!)^{a_2}a_2!\cdots((k-1)!)^{a_{k-1}}a_{k-1}!}
(\phi)^{a_1}(\phi')^{a_2}\ldots(\phi^{(k-2)})^{a_{k-1}} \;\;,
\end{equation}
where $\sum_{\{\vec{a}_{k-1};m\}}$ stands for the summation with
$a_1,a_2,\ldots,a_{k-1}$ running over all positive integers under the
restrictions that
\[
a_1+2 a_2+ 3 a_3 +\ldots+(k-1)a_{k-1}=k-1 \qquad\textrm{and}\qquad
a_1+a_2+\ldots+a_{k-1}=m \;\;.
\]
This sum can be interpreted following the time-honored formula of {\em Fa\`a
di Bruno}~\cite{Abramowitz}: 
\[
\frac{d^n}{dx^n}f(\,g(x)\,)=\sum_{m=0}^{n} f^{(m)}(\,g(x)\,)
\sum_{\{\vec{a}_{n};m\}}(n;a_1,\ldots,a_n)
\{g'(x)\}^{a_1} \{g''(x)\}^{a_2}\cdots\{g^{(n)}(x)\}^{a_n}
\]
where 
\[ (n;a_1,\ldots,a_n)=\frac{n!}{
(1!)^{a_1}a_1!(2!)^{a_2}a_2!\cdots(n!)^{a_n}a_n!} \;\;.
\]
The identifications
$g'(x)=\phi(x)$ and $f^{(m)}(\,g(x)\,)=\hbar^{-m}f(\,g(x)\,)$
with the solution
\begin{equation}
g(x)=\int dx \,\phi(x)\quad,\qquad
f(\,g(x)\,)=R(x)=\exp\left(\frac{1}{\hbar}\int dx\, \phi(x)\right) \;\;,
\end{equation}
lead to an equation for $R$, which, including 
all possible vertices, reads
\begin{equation}
  \sum_{k=3}^{\infty}\frac{\lambda_k}{(k-1)!}\,\hbar^{k-1}R^{(k-1)}
  + \mm\hbar R'(x) - (x-\lambda_1) R(x) 
  = 0\;\;.
\label{LinSD}  
\end{equation}
This is a linear equation, and a
solution can be represented by an integral
\begin{equation}
   R_{\Gamma}(x) = \int_{\Gamma}d\vhi\,
   \exp\left\{\frac{1}{\hbar}[x\vhi-S(\vhi)]\right\} \;\;,
\label{gpath}				   
\end{equation}
where
\begin{equation}
   S(\vhi)=\lambda_1\vhi+\frac{1}{2}\mm\vhi^2
+\sum_{k=3}^{\infty}\frac{\lambda_k}{k!}\,\vhi^k
   \;\;,
\nn\end{equation}
and where $\Gamma$ is a contour in the complex $\vhi$-plane, such that the
difference between the values of the integrand in the end-points is zero. This
is the well known {\em path integral representation}, with the {\em action}
$S$.

A remark is in order. Although the \SD\ equations resulted from a a
purely diagrammatic construction, their solutions, expressed through the path
integral representation, include non-perturbative ones that cannot be  
realized
in a weak coupling expansion, as we will see below.

Secondly, we note that \eqn{LinSD} can be used to write the original \SD\ 
equation for $\phi$ compactly as
\begin{equation}
   x = \la_1 + \mu\phi 
       + \sum_{k\geq3}\frac{\la_k}{(k-1)!}
         \left(\hbar\frac{\pa}{\pa x} + \phi\right)^{k-2}\phi \;\;.
\end{equation}
For a general interacting theory, differantiating $R_\Gamma$ with 
respect to $\lambda_k$ in \eqn{gpath}, the stepping equation in terms
of $\phi$ can be rewritten as
\begin{equation}
\frac{\partial\phi}{\partial \lambda_k}=
-\frac{1}{k!}\frac{\partial}{\partial x}\left(\phi+\hbar
\frac{\partial}{\partial x}
\right)^{k-1}\phi
\end{equation}
and in case only a $k$-vertex and the tadpole $\lambda_1$ is present, 
combining with the \SD\ a simpler form is obtained  
\begin{equation}
\frac{\partial\phi}{\partial \lambda_k}=
-\frac{1}{k\lambda_k}
\big(\left(x-\lambda_1\right)\phi'+\phi-2\mu\phi\phi'-\hbar\mu\phi''
\big)\;.
\label{newsteppingeq}\end{equation}
Moreover for a charged scalar field the stepping equation 
in terms of $\phi$ can be written in a compact form
\begin{equation}
\frac{\partial\phi}{\partial\lambda}=
-\frac{1}{4}\bar{\partial}(\bar{\phi}+\hbar\partial)^2
(\phi+\hbar\bar{\partial})\phi\;.
\end{equation}

Finally, the linear \SD\ equation for $\vhi^3+\vhi^4$-theory becomes simply 
\begin{equation}
  \sixt\laf\hbar^3R'''(x) + \half\lat\hbar^2R''(x) + \mm\hbar R'(x) - xR(x) 
  = 0\;\;,
\label{rsd}
\end{equation}
and we see that $R(x)$ admits 3 linearly independent solutions
(2 if $\laf=0$). Hence $\phi(x)$ has a 2-parameter family of solutions
(a 1-parameter family if $\laf=0$). 
In the sequel we will
show how to get exact explicit solutions for a number of scalar theories.

%
%
\subsection[Results for pure ${\vhi^3}$-theory]%
           {Results for pure $\boldsymbol{\vhi^3}$-theory\label{Sec3}}
In this section we derive results for the pure $\vhi^3$-theory, with action
\begin{equation}
   S(\vhi) = \frac{1}{2}\mm\vhi^2 + \frac{1}{6}\lambda\vhi^3
   \;\;.
\end{equation}
This theory is interesting because as we will see the solution for the
generating function can be expressed directly in terms of known special
functions.  Defining 
\[   y = \frac{x}{\sqrt{\hbar\mm}} \quad,\qquad
   \xi = \frac{\lambda \sqrt{\hbar}}{6 \mm^{3/2}}
\]
the \SD\ equation becomes
\begin{equation}
   3\xi R''(y)+R'(y)-yR(y)=0
\label{ZeEq003}   
\end{equation}
which admits the following general solution
\begin{equation}
   R(y)=e^{-y/6\xi}\left[ c_1 \Ai(t) +c_2 \Bi(t) \right]
\label{R-solution}\end{equation}
where
\[ t= (3\xi)^{-1/3}\left(\frac{1}{12 \xi}+y\right). \]
$\Ai$ and $\Bi$ are the Airy functions (cf. \cite{Abramowitz}). 
The solution for the generating function of connected Green's functions is
given by
\begin{equation}
   \phi(x) = \sqrt{\frac{\hbar}{\mm}}\left( -\frac{1}{6\xi}+ 2^{1/2}  t_0^{1/4} 
             \frac{\Ai'(t)+K\Bi'(t)}{\Ai(t)+K\Bi(t)} \right)
\end{equation}
with $t_0=t(y=0)$. The constant $K$ is not determined by the \SD\ equation: in
fact it could have been even a function of $\xi$.

For solutions that admit a diagrammatic representation extra information
can be obtained by combining \SD\ and stepping equations.
For instance, the scaling 
and stepping equations of the previous section result to a $K$ that is 
independent of $\xi$.
Moreover by combining \SD\  and \Stw\ an equation involving only
$\C{1}$:
\begin{equation}
   2\mm^2\C{1} + \lambda\mm \C{1}^2 + \hbar\lambda^2\ddmu \C{1} 
   + \hbar\lambda = 0\;\;,
\label{c1phi3}\end{equation}
can be obtained.
The series of substitutions
\begin{equation}
   v = \frac{\hbar\lambda^2}{\mm^3}\;,\;\;
   \C{1} = -\frac{\lambda\hbar}{2\mm^2}f(v)\;,\;\;
   w = \frac{1}{3v}\;,\;\;
   f(v) = -\frac{2k'(w)}{ vk(w)}\;,\;\;
   k(w) = w^{1/3}e^{-w}\psi(w)\;\;,
\nn\end{equation}
leads to the Bessel equation
\begin{equation}
   w^2\psi''(w) + w\psi'(w) - \left(w^2+\frac{1}{9}\right)\psi(w) = 0\;\;.
\nn\end{equation}
The special solution choice $\psi(w)=K_{1/3}(w)$ gives the following
tadpole and its asymptotic expansion:
\begin{align}
   f(v) &= \frac{2}{v}\left(\frac{K_{2/3}(w)}{K_{1/3}(w)}+1\right)\;\;,\nl
   \C{1} &\sim -2\frac{\mm}{\lambda}
   -\frac{\hbar\lambda^2}{2\mm^2}\left
   (1 - \frac{5}{4}v + \frac{15}{4}v^2 - \frac{1105}{64}v^3 + 
   \frac{1695}{16}v^4 + \cdots \right)\;\;.
\label{c1phi3expansion}\end{align}
This tadpole, therefore, has a non-perturbative contribution. The more
generic choice
$\psi=k_1I_{1/3}(w)+k_2I_{-1/3}(w)$, with $k_1\ne-k_2$, gives
\begin{align}
   f(v) &= -\frac{2}{v}\left(\frac{k_1I_{-2/3}(w)+k_2I_{2/3}(w)}
                              { k_1I_{1/3}(w)+k_2I_{-1/3}(w)} -1\right)\;\;,\nl
   \C{1} &\sim -\frac{\hbar\lambda^2}{2\mm^2}\left
   (1 + \frac{5}{4}v + \frac{15}{4}v^2 + \frac{1105}{64}v^3 + 
   \frac{1695}{16}v^4 + \cdots \right)\;\;,
\label{c1phi3expansiona}\end{align}
which is the standard perturbative result~\cite{Cvitanovic}. 
The coefficients $k_1$ and $k_2$ 
drop out for the perturbative expansion: they simply account for 
non-perturbative contributions that are not computable perturbatively!

A remark is in order here: although all the terms in the perturbative series
for $\C{1}$ have strictly the same complex phase and according to the
traditional wisdom the series is not Borel summable, the exact result is well
defined, indicating that a suitable generalization of the Borel transform will
produce the right answer~\cite{Ellis}.

Another interesting aspect is the large $n$ behavior of the Green's functions,
where $n$ refers to the number of external legs, a problem that is
traditionally seen as relevant to the unitarity of the $S$
matrix~\cite{Goldberg}. This can be traced from the analytical structure of
the solution for the generating functions in the complex $x$-plane. As is
evident from the fact that the solution, \eqn{R-solution}, for the generating
function of all connected and disconnected graphs is an entire function,
the corresponding Green's function $Z_n$ grows slower than $n!$; in fact it
grows like $(n!)^{2/3}$. On the other hand the $\C{n}$, the connected graphs,
exhibit a factorial growth, since their generating function $\phi(x)$ posesses
poles at finite complex values of $x$.

%
%
\subsection[Results for pure ${\vhi^4}$-theory]%
           {Results for pure $\boldsymbol{\vhi^4}$-theory\label{Sec4}}
In this section we derive the lowest Green's functions for the pure $\vhi^4$
theory, with action
\begin{equation}
   S(\vhi) = \frac{1}{2}\mm\vhi^2 + \frac{1}{24}\lambda\vhi^4
   \;\;.
\end{equation}
Defining 
\[ y=\frac{x}{\sqrt{\hbar\mm}} \quad,\qquad \xi=\frac{\lambda 
\sqrt{\hbar}}{24 \mm^2}
\]
we get for the \SD\ equation
\begin{equation} 
    4\xi R'''(y) + R'(y) - yR(y) = 0\;\;.
\label{SoEq439}
\end{equation}
There are three solutions, which can be represented as follows:
\begin{align}
   R_1(y) &= \sum_{n=0}^\infty \frac{y^{2n}}{n!}\left(32\xi\right)^{-n/2}
   \mbox{U}(n;(8\xi)^{-1/2}) 
\nl
   R_2(y) &= \sum_{n=0}^\infty (-1)^n\frac{y^{2n}}{n!}\left(32\xi\right)^{-n/2}
   \frac{\mbox{V}(n;(8\xi)^{-1/2})}{\Gamma(n+\frac{1}{2})} 
\nl
   R_3(y) &= \sum_{n=0}^\infty \frac{y^{2n+1}}{(2n+1)!}
   \left(4\xi\right)^{-n/2}\,i^n\mbox{H}_n(i(16\xi)^{-1/2}) \;\;,
\label{SoEq440}
\end{align}
where $\mbox{U}(\nu;x)$ and $\mbox{V}(\nu;x)$ are the parabolic cylinder
functions, and $\mbox{H}_n$ is the $n^{\textrm{th}}$ Hermite polynomial 
(cf. \cite{Abramowitz}).
The general solution is a linear combination with arbitrary coefficients.  As
we can immediately see contrary to what is argued in many standard textbooks
the odd Green's functions do not necessarily vanish. 

On the other hand on can study the \Stw\ equation as well. For this we have to
distinguish two possible cases: the `standard' one, with $\C{1}$ and the higher
odd Green's functions vanishing, and the case where $\C{1}\ne 0$.

Let us consider the first case with a zero tadpole.
In this case we cannot, of course, directly use the results derived above,
since these deal with $\C{1}$.  The \Stw\ becomes somewhat simpler, and in
particular
\begin{equation}
   \C{4} = -\frac{2}{\hbar}\left(\C{2}^2 + \ddmu \C{2}\right)\;\;.
\nn\end{equation}
On dimensional grounds we see that we can write
\begin{equation}
   \C{2} = \frac{1}{\mm}\beta(v) \quad,\qquad
   v=\frac{\lambda\hbar}{\mm^2}\;\;,
\nn\end{equation}
where $v$ is dimensionless.
Inserting all this into the first nonzero term (that with $x^1$) in \SD, we
find the following equation for $\beta$:
\begin{equation}
   4v^2\beta'(v) + v\beta(v)^2 + (2v+6)\beta(v) - 6 = 0\;\;.
\nn\end{equation}
The substitutions
\begin{equation}
   \beta(v) = 4v\frac{g'(v)}{ g(v)} \quad,\qquad
   g(v) = v^{-1/4}e^w\psi(w) \quad,\qquad 
   w=\frac{3}{4v} \;\;,
\label{betaeqn}
\end{equation}
lead then to
\begin{equation}
   w^2\psi''(w) + w\psi'(w) - \left(w^2+\frac{1}{16}\right)\psi(w) = 0\;\;,
\label{besseleqn}
\end{equation}
which has the modified Bessel functions for its solutions. 
The general solution can always be written as 
\begin{equation}
   \psi(w) = k_1I_{1/4}(w) + k_2I_{-1/4}(w)\;\;.
\nn\end{equation}
It is instructive to consider the perturbative form of these results, that is,
the limit where $\hbar$ becomes infinitesimally small, or $w$ goes to
infinity.  Since $I_{1/4}$ and $I_{-1/4}$ have the same asymptotic expansion,
a generic choice of $k_{1,2}$ will lead to a single perturbative expansion.
The single exception is the choice $k_1=-k_2$ which leads to $\psi(w)\propto
K_{1/4}(w)$, with asymptotic expansion
\begin{equation}
\beta(v) = \frac{3}{v}\left(\frac{K_{3/4}(w)}{ K_{1/4}(w)} - 1\right)
\sim 1 - \frac{1}{2}v + \frac{2}{3}v^2 - \frac{11}{8}v^3 
       + \frac{34}{9}v^4 + \cdots\;\;.
\nn\end{equation}
This is the standard perturbative  expansion, in which the propagator starts
with $\frac{1}{\mm}$, and has loop corrections in powers of $\hbar$:
\begin{equation}
\C{2} = \C{2}^{(1)} = \frac{1}{\mm}\left( 1 - \frac{\lambda\hbar}{2\mm^2}
+ \frac{2\lambda^2\hbar^2}{3\mm^4} + \cdots\right)\;\;.
\end{equation}
The alternating signs are of course due to the fact that the Feynman rules
prescribe a factor $-\lambda$ for each vertex in our Euclidean model. The
asymptotic expansion in all other cases is equal to that for the choice 
$k_2=0$, for which we find
\begin{equation}
\beta(v) = -\frac{3}{v}\left(
\frac{I_{-3/4}(w)}{ I_{1/4}(w)} + 1\right)
\sim -\frac{6}{ v} + 1 + \frac{1}{2}v + \frac{2}{3}v^2 
                    + \frac{11}{8}v^3 + \frac{34}{9}v^4 + \cdots\;\;,
\nn\end{equation}
which gives a nonstandard expansion:
\begin{equation}
\C{2} = \C{2}^{(2)} = -\frac{6\mm}{\lambda\hbar} + 
\frac{1}{\mm}\left( 1 + \frac{\lambda\hbar}{2\mm^2}
+ \frac{2\lambda^2\hbar^2}{3\mm^4} + \cdots\right)\;\;.
\end{equation}
Note the occurrence of a `non-perturbative' term $1/\lambda$ here: the rest of
the expansion has an apparent opposite sign of the coupling constant.
An other way to look at this solution is by examining the
saddle point equation, $\delta S/\delta\phi =x$: the abovementioned 
solution corresponds to the saddle point $\phi_c=\sqrt{- 6\mm/\lambda}
+{\cal O}(x)$.

In the case $C_1\ne 0$ we can write, again on dimensional grounds,
\begin{equation}
   \C{1} = \alpha(v)\sqrt{\frac{\mm}{\lambda}} \quad,\qquad
   \C{2} = \frac{1}{\mm}\beta(v)\;\;,
\nn\end{equation}
with $v$ as before.  The first term (with $x^0$) in \SD\ now gives us a
relation between $\alpha$ and $\beta$:
\begin{equation}
   \beta(v) = \frac{1}{ v\alpha(v)}\left(
   (6-v)\alpha(v) + 4v^2\alpha'(v)\right)\;\;,
\nn\end{equation}
and then the second term ($x^1$) gives
\begin{equation}
    16v^2\alpha(v)\alpha''(v) - 32v^2\alpha'(v)^2 
    + (32v-24)\alpha(v)\alpha'(v) - 3\alpha(v)^2 = 0\;\;.
\nn\end{equation}
Using $w$ as before, we may now substitute
\begin{equation}
   \alpha(v) = \frac{e^w\sqrt{v}}{\psi(w)}\;\;,
\nn\end{equation}
to find that $\psi(w)$ again obeys the Bessel equation, \eqn{besseleqn}.
For the asymptotic expansions, again two distinct choices are possible.
First, the choice
\begin{equation}
   \psi(w) = \frac{1}{ p}K_{1/4}(w)
\nn\end{equation}
gives
\begin{equation}
   \alpha(v) = \frac{pe^w\sqrt{v}}{K_{1/4}(w)} \quad,\qquad
   \beta(v) = \frac{3}{v}\left(\frac{K_{3/4}(w)}{K_{1/4}(w)} - 1\right)
              - \frac{p^2e^{2w}}{K_{1/4}(w)^2}\;\;,
\nn\end{equation}
and the following asymptotic forms for $\C{1,2}$:
\begin{equation}
   \C{1} \sim p\sqrt{\frac{3\mm}{2\pi\lambda}}e^{2w} \quad,\qquad
   \C{2} \sim \C{2}^{(1)} + \frac{2p^2we^{2w}}{\mm\pi}\;\;.
\end{equation}
The alternative choice, for which we may take
\begin{equation}
   \psi(w) = \frac{1}{ p}I_{1/4}(w)\;\;,
\nn\end{equation}
leads to
\begin{equation}
   \alpha(v) = \frac{pe^w\sqrt{v}}{ I_{1/4}(w)} \quad,\qquad
   \beta(v) = -\frac{3}{v}\left(\frac{I_{-3/4}(w)}{ I_{1/4}(w)} + 1\right)
              - \frac{p^2e^{2w}}{ I_{1/4}(w)^2}\;\;,
\nn\end{equation}
and
\begin{equation}
   \C{1} \sim p\sqrt\frac{3\pi\mm}{2\lambda} \quad,\qquad
   \C{2} \sim \C{2}^{(2)} 
              - \frac{2\pi p^2v}{\mm}\left(1 - \frac{1}{4}v -\frac{13}{96}v^2 
	                             - \frac{73}{384}v^3 - \cdots\right)\;\;.
\end{equation}
In contrast to the zero-tadpole case, there remains an arbitrary parameter
in these solutions, $p$: it reflects the presence of the `non-perturbative' 
tadpole-like contribution and has to be determined by additional 
requirements.

%
%
\subsection[Results for $\vhitf$-theory]
           {Results for $\boldsymbol{\vhitf}$-theory}
For the general zero-dimensional $\vhitf$-theory, the action is given by
\begin{equation}
   S(\vhi) = \half\mm\vhi^2 + \sixt\lat\vhi^3 + \twft\laf\vhi^4 \;\;.
\end{equation}
In the dimensionless variables
\begin{equation}
   y = \frac{x}{\sqrt{\mm\hbar}} \;\;,\quad
   \gt = \frac{\lat}{\mm}\sqrt{\frac{\hbar}{\mm}} \;\;,\quad
   \gf = \frac{\laf\hbar}{\mm^2}  \;\;,
\nn\end{equation}
the \SD\ equation becomes
\begin{equation}
   \sixt\gf R'''(y) + \half\gt R''(y) + R'(y) - yR(y) = 0 \;\;.
\label{SoEq341}   
\end{equation}
To solve this equation, let
\begin{equation}
   R(y) = e^{-y\gt/\gf}F(y) \;\;.
\nn\end{equation}
Then $F$ satisfies the equation
\begin{equation}
   \sixt\gf F'''(y) + \alpha F'(y) - (y+\beta)F(y) = 0 \;\;,
\label{SoEq342}   
\end{equation}
where
\begin{equation}
   \alpha = 1-\frac{\gt^2}{2\gf} \qquad,\qquad
   \beta = \frac{\gt}{\gf}\left(1-\frac{\gt^2}{3\gf}\right)  \;\;.
\nn\end{equation}
Finally, changing variables 
\begin{equation}
   y+\beta = \frac{\eta}{\sqrt{\alpha}} \qquad,\qquad
   4\xi = \frac{\gf}{6\alpha^2} 
        = \frac{\gf}{6}\left(1-\frac{\gt^2}{2\gf}\right)^{-2} \;\;,
\nn\end{equation}
\eqn{SoEq342} becomes 
\begin{equation}
   4\xi F'''(\eta) + F'(\eta) - \eta F(\eta) = 0 \;\;.
\label{SoEq343}   
\end{equation}
\eqn{SoEq343} is exactly \eqn{SoEq439} of the pure $\vhi^4$-theory, so 
that the solutions here are those of (\ref{SoEq440}) with $\xi$ as given 
above and $y$ replaced by $\eta$.

%
%
\subsection{Results for the charged scalar field}

For the  complex scalar field, the path integral solution is given by  
\begin{equation}
   R(\xx,\xb) = \int d\vhi d\vhib\,\exp\left\{\frac{1}{\hbar}
       [\xx\vhib+\xb\vhi-S(\vhi,\vhib)]\right\} \quad,\quad
  S(\vhi,\vhib)=\mm\vhib\vhi+\frac{\la}{4} \left(\vhib\vhi\right)^2 \;\;.
\label{ChScAc}
\end{equation}
Due to charge conservation ($\mbox{O}(2)$-symmetry) one can easily 
show that $R$ only depends on the modulus $\xx\xb$, so that it satisfies the
following equation
\begin{equation}
   \zeta R'''(\zeta)+2 R''(\zeta) +\alpha^2[R'(\zeta)-R(\zeta)]=0 \quad,\qquad
   \zeta = \frac{\xx\xb}{\mm\hbar}  \;\;,\quad
   \alpha = \mm\left(\frac{2}{g\hbar}\right)^{1/2} \;\;.
\label{SoEq064}   
\end{equation}
The third order equation can be solved by power series expansion in $\zeta$
and two of its solutions are given by
\begin{equation}
   R_1(\zeta) = \sum_{n=0}^\infty\frac{1}{i(n!)^2}
                \left(\frac{i\zeta \alpha}{\sqrt{2}}\right)^n 
                \mbox{H}_n\!\left(\frac{i\alpha}{\sqrt{2}}\right) \;\;,\quad
  R_2(\zeta) = \sum_{n=0}^\infty\frac{(\zeta \alpha)^n}{n!}\,
               \mbox{U}(n+{\textstyle\frac{1}{2}},\alpha)          \;\;,
\end{equation}
where $\mbox{H}_n$ stands for the $n^{\textrm{th}}$ order Hermite polynomial
and $\mbox{U}(\nu,x)$ is the parabolic cylinder function.
The third solution can be found by standard procedures but the
actual result is rather cumbersome and we refrain from giving it
explicitly.

Where $R_1$ is a purely non-perturbative solution, $R_2$ has
an asymptotic series expansion which leads to the normal
perturbation series for the connected Green's functions:
for instance the two point function is given by
\begin{equation}
   \C{1,1} = \frac{1}{\mm} - \frac{g\hbar}{\mm^3} 
             + \frac{5}{2}\frac{g^2\hbar^2}{\mm^5} + \mathcal{O}(g^3\hbar^3)
	     \;\;.
\end{equation}
Although the standard integral representation of $R$ is given by 
(\ref{ChScAc}), 
the differential equation in the variable $\zeta$ leads to another
peculiar single contour integral representation 
\[
   R(\zeta) = \int_{\Gamma} d\psi\,\exp\left(\zeta\psi-\ln\psi 
              -\frac{\alpha^2}{\psi}+\frac{\alpha^2}{2\psi^2}\right) \;\;,
\]
where the contour $\Gamma$ is from infinity to infinity such that the integral
is convergent. 

%
%
\subsection{Contours in the integral representation}
\begin{figure}
\begin{center}
\epsfig{figure=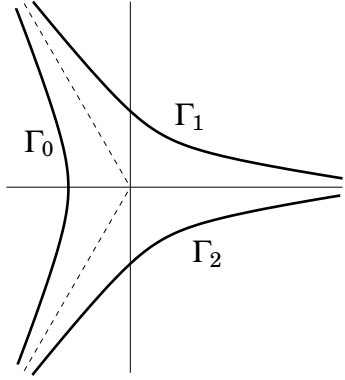,width=0.28\linewidth}
\caption{Contours in the complex $u$-plane for the Airy functions.}
\label{FigAiryCont}
\end{center}
\end{figure}

The integral representation of the solutions for the
pure $\vhi^3$-theory, with 
$\psi= \sqrt{\mm/\hbar}\,\vhi$, can be written as 
%
\begin{equation}
   R(y;\xi) = K\int_{\Gamma}
              d\psi\,\exp\left(-\frac{1}{2}\psi^2-\xi\psi^3+y\psi\right) \;\;,
\end{equation}
where $K$ is a constant which can depend on $\xi$.
In case the moduli of the endpoints of the contour $\Gamma$ are taken to 
infinity the standard path-integral representation is recovered.
In fact in this case the substitution 
\begin{equation}
   u=(3\xi)^{-1/3}\psi-\frac{1}{6\xi}
\nn\end{equation}
leads to
\begin{equation}
   R(y;\xi) = K \exp\left(-\frac{1}{108\xi}-\frac{y}{6\xi}\right)
              \int_\Gamma du\,\exp\left(-\frac{1}{3}u^3+tu\right) \;\;.
\nn\end{equation}
This integral can now be expressed in terms of the Airy functions
\begin{equation}
   \int_{\Gamma_j} du\,\exp\left(-\frac{1}{3}u^3+tu\right) 
   = 2\pi\omega^j \Ai(t \omega^j)  \;\;,
\nn\end{equation}
where $\omega=e^{i2\pi/3}$,
$j=0,1,2$ and the contours $\Gamma_j$ are depicted in \fig{FigAiryCont}.
Note that 
\[ \sum_{j} \omega^j\Ai(t\omega^j)=0 \;\;.
\]
For a pure $\vhi^4$-theory, similar considerations allow us to 
express the functions $R_j$ defined in \eqn{SoEq440}, $j=1,2,3$ as follows
\begin{align}
   R_1 &= \frac{1}{\sqrt{\pi}}(2\xi)^{1/4}\exp\left(-\frac{1}{32\xi}\right)
   \int_{-\infty}^\infty d\psi\,
   \exp\left(-\frac{1}{2}\psi^2-\xi\psi^4+y\psi\right)
\nl
   R_2 &= \frac{-i}{\pi^{3/2}}(2\xi)^{1/4}\exp\left(-\frac{1}{32\xi}\right)
   \int_{-i\infty}^{i\infty} d\psi\,
   \exp\left(-\frac{1}{2}\psi^2-\xi\psi^4+y\psi\right)
\end{align}
whereas for $R_3$  
\begin{multline}
  \left( \int_0^\infty+\int_0^{-i\infty} \right)
  d\psi\,\exp\left(-\frac{1}{2}\psi^2-\xi\psi^4+y\psi\right) \\
  = \frac{\sqrt{\pi}}{2}(2\xi)^{-1/4}\exp\left(\frac{1}{32\xi}\right)
    (R_1-i \pi R_2) + \frac{i\pi}{2}\exp\left(\frac{1}{16\xi}\right) 
                      (2\xi)^{-1/4}R_3 \;\;.
\end{multline}

\begin{figure}
\begin{center}
\epsfig{figure=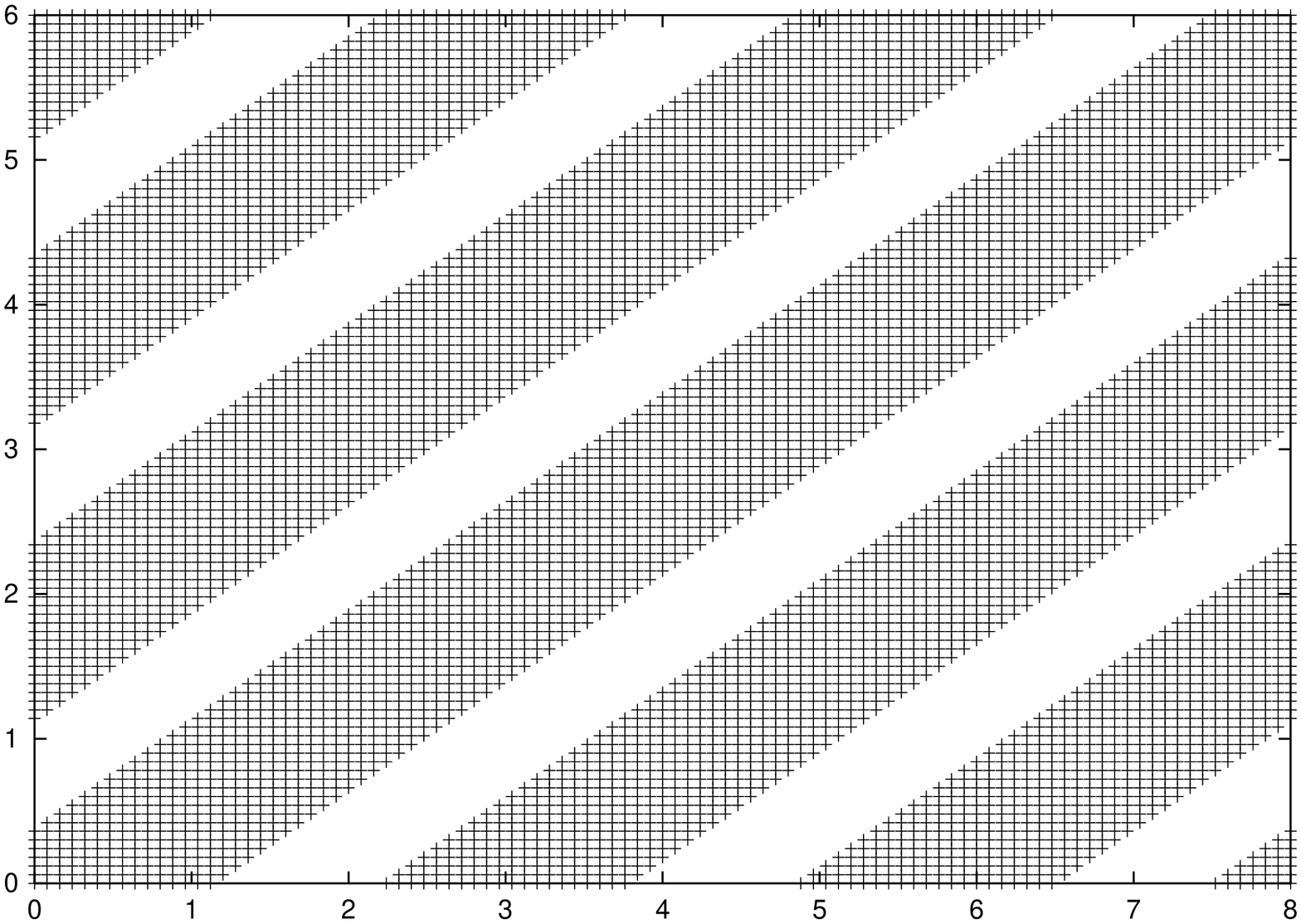,width=0.33\linewidth,angle=270}
\epsfig{figure=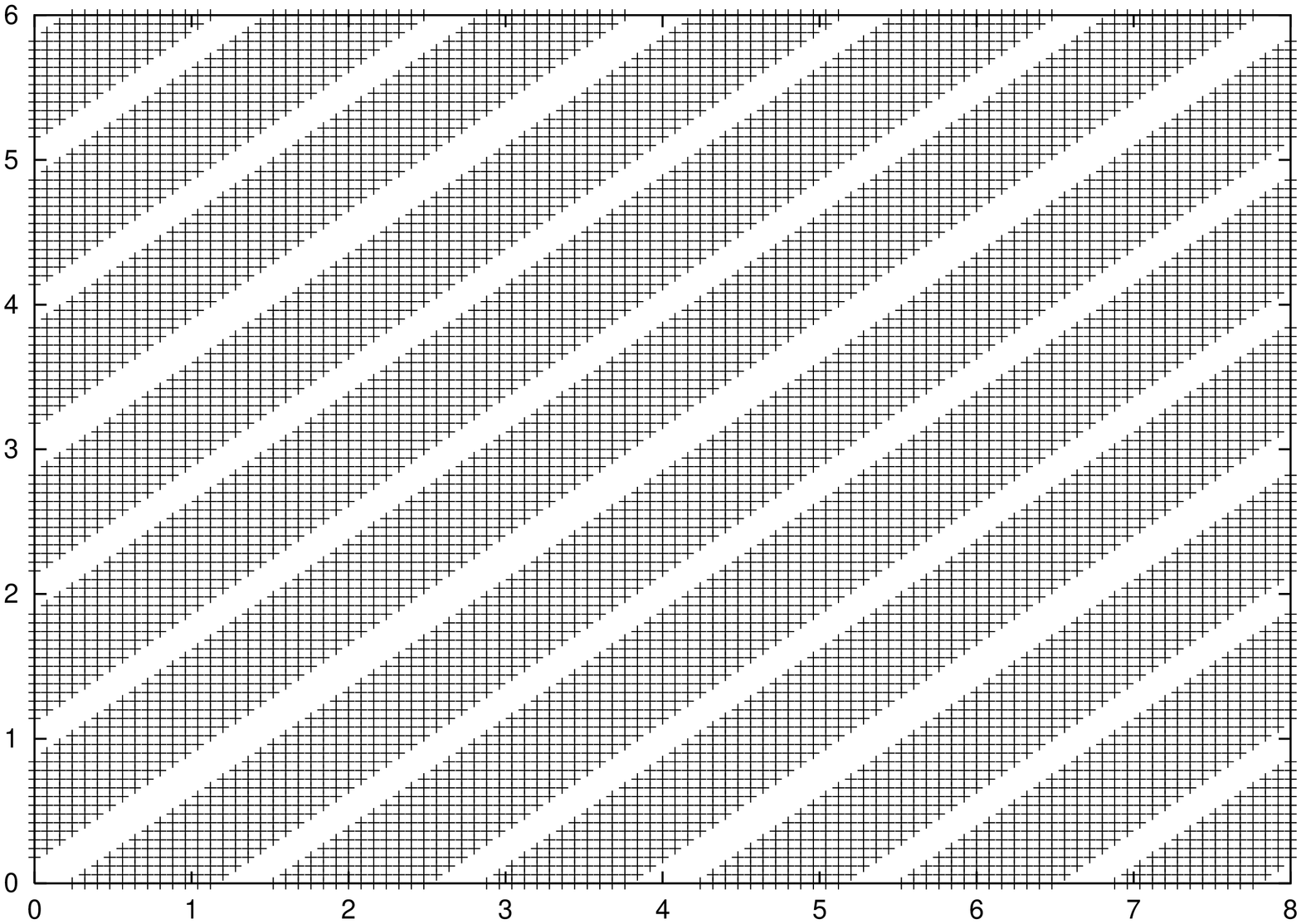,width=0.33\linewidth,angle=270}
\caption[.]{The shaded areas correspond to combinations of $\eta_4$ and 
            $\eta_3$ (in units of $\pi$) for which the limit $\laf\to0$ 
	    (with $\lat$ fixed) is perturbative, for contour $\Gamma_{01}$ 
	    (left plot) and contour $\Gamma_{02}$ (right).}
\label{shadplot1}
\end{center}
\end{figure}

\begin{figure}
\begin{center}
\epsfig{figure=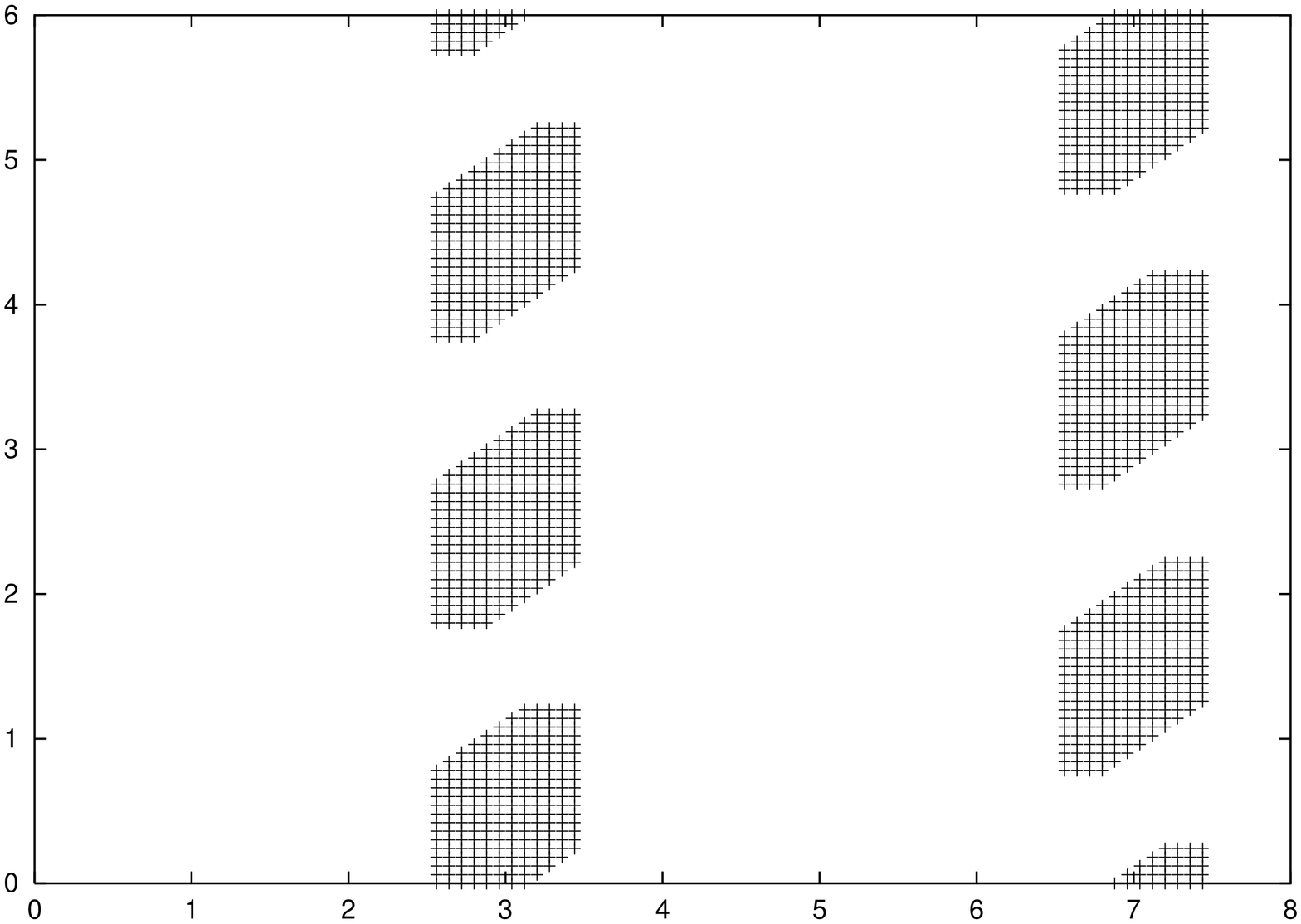,width=0.33\linewidth,angle=270}
\epsfig{figure=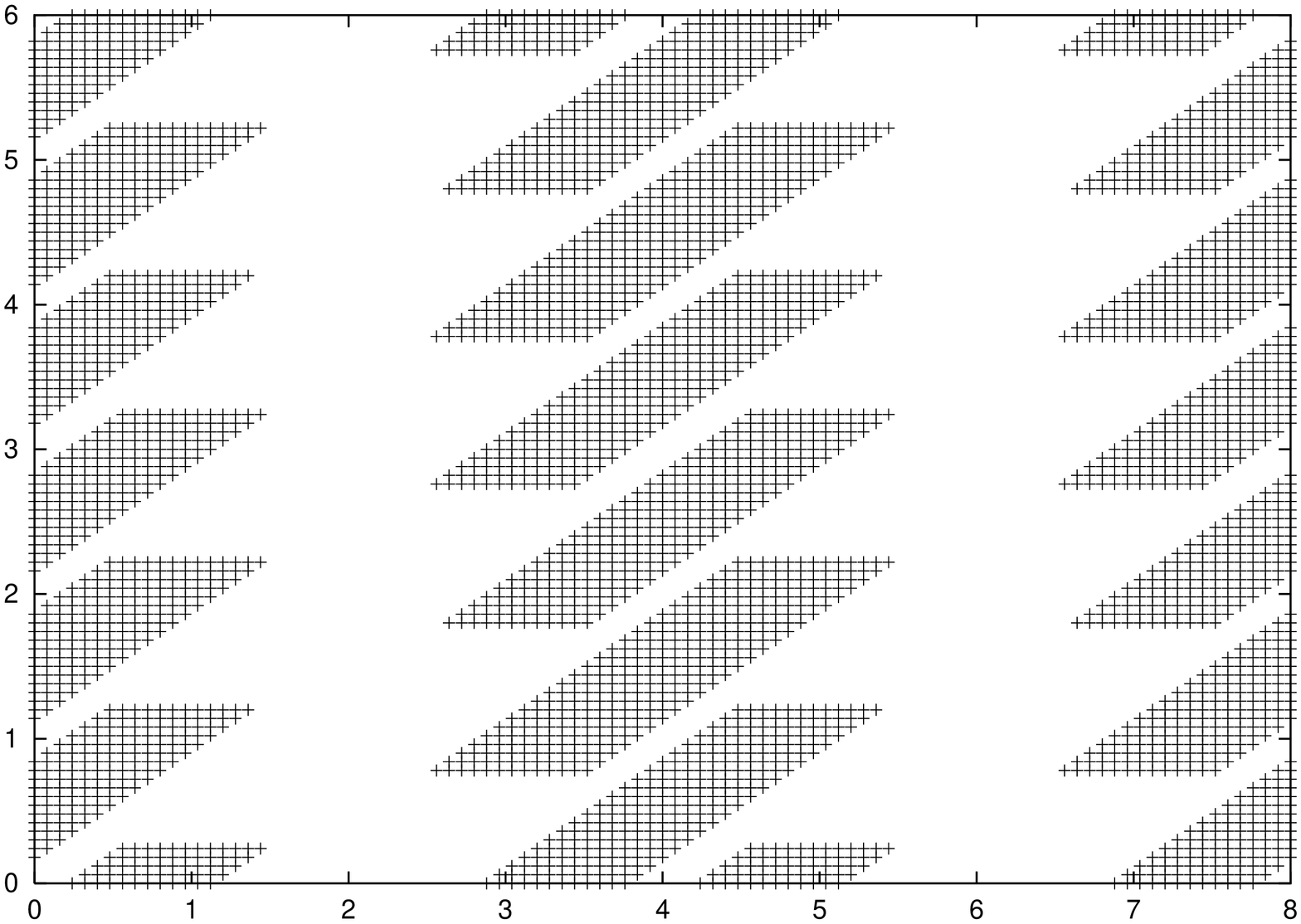,width=0.33\linewidth,angle=270}
\caption[.]{The shaded areas correspond to combinations of $\eta_4$ and 
            $\eta_3$ (in units of $\pi$) for which the limit $\laf\to0$ 
	    followed by $\lat\to0$ is perturbative, for contour $\Gamma_{01}$ 
	    (left) and $\Gamma_{02}$.}
\label{shadplot2}
\end{center}
\end{figure}

Let us have a closer look at the various possible contours in the case of
general $\vhitf$-theory.  Let us denote the various objects in the action
as complex numbers:
\begin{equation}
\vhi = |\vhi|e^{i\omega}\;\;\;,\;\;\;
\lat = |\lat|e^{i\eta_3}\;\;\;,\;\;\;\laf = |\laf|e^{i\eta_4}\;\;.
\nn\end{equation}
For simplicity and without loss of generality, we may keep $\mm$ real 
and positive.
The direction in the $\vhi$ plane where the term $\laf\vhi^4$ goes to positive infinity
as $|\vhi|\to\infty$ are given by
\begin{equation}
\omega \in \Omega_4(k)\;\;\;,\;\;\;
\Omega_4(k) = \left( k\frac{\pi}{2} - \frac{\pi}{8} - \frac{\eta_4}{4}\;,\;
k\frac{\pi}{2} + 
\frac{\pi}{8} - \frac{\eta_4}{4} \right)\;\;,\;\;k=0,1,2,3\;\;.
\nn\end{equation}
Similarly `allowed' directions for the $\lat\vhi^3$ term are
\begin{equation}
\omega \in \Omega_3(k)\;\;\;,\;\;\;
\Omega_3(k) = \left( k\frac{2\pi}{3} - \frac{\pi}{6} - \frac{\eta_3}{3}\;,\;
k\frac{2\pi}{3} + \frac{\pi}{6} - \frac{\eta_3}{3} \right)\;\;,\;\;k=0,1,2\;\;.
\nn\end{equation}
Finally, the $\mm\vhi^2$ term goes to positive infinity for
\begin{equation}
\omega \in \Omega_2(k)\;\;\;,\;\;\;
\Omega_2(k) = \left( k\pi - \frac{\pi}{4}\;,\;k\pi +\frac{\pi}{4} \right)\;\;,k=0,1\;\;.
\nn\end{equation}
By inspection of these endpoints, already statements can be made about the
(non)per\-tur\-bative character of the theory corresponding to a given contour.
\begin{figure}
\begin{center}
\epsfig{figure=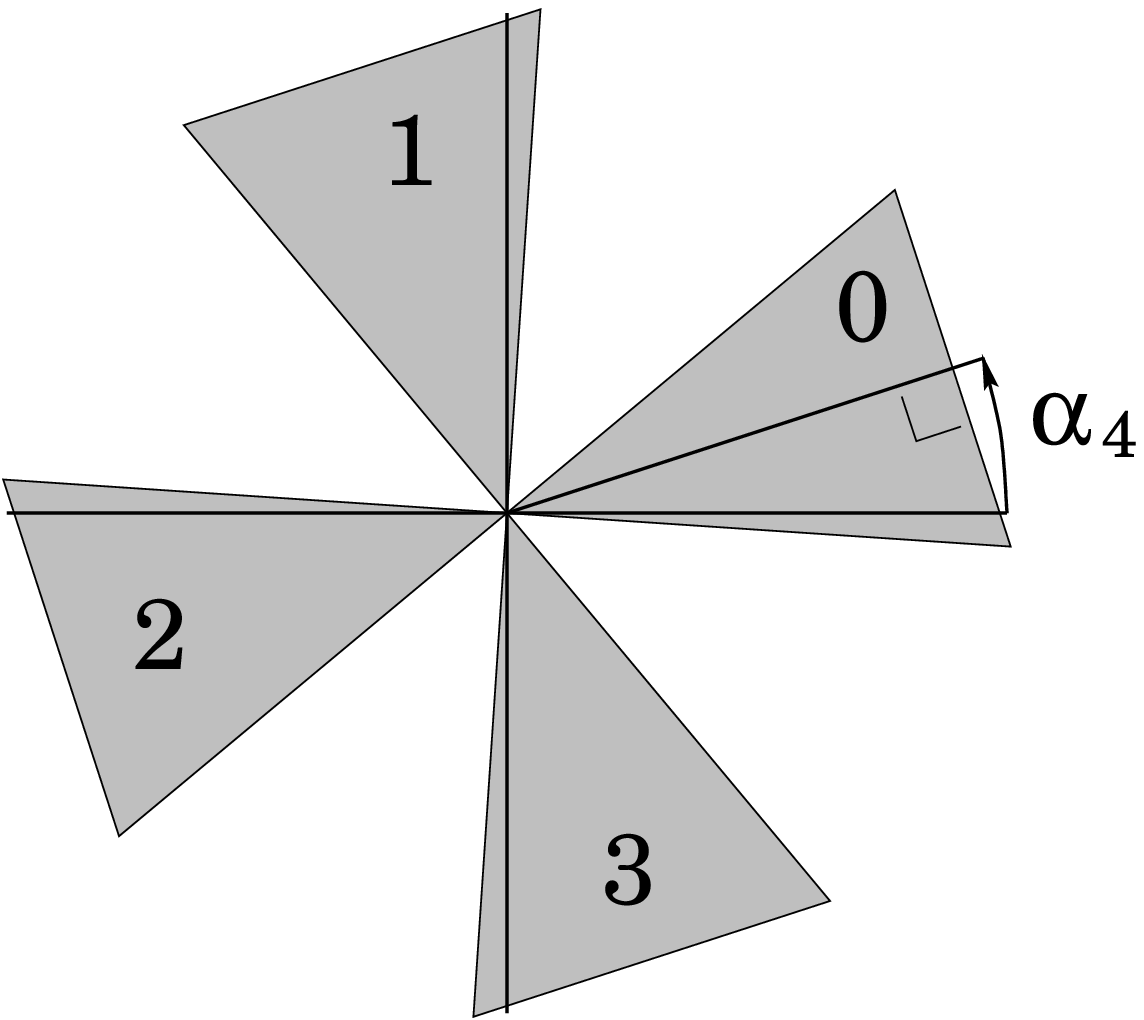,width=0.31\linewidth}
\epsfig{figure=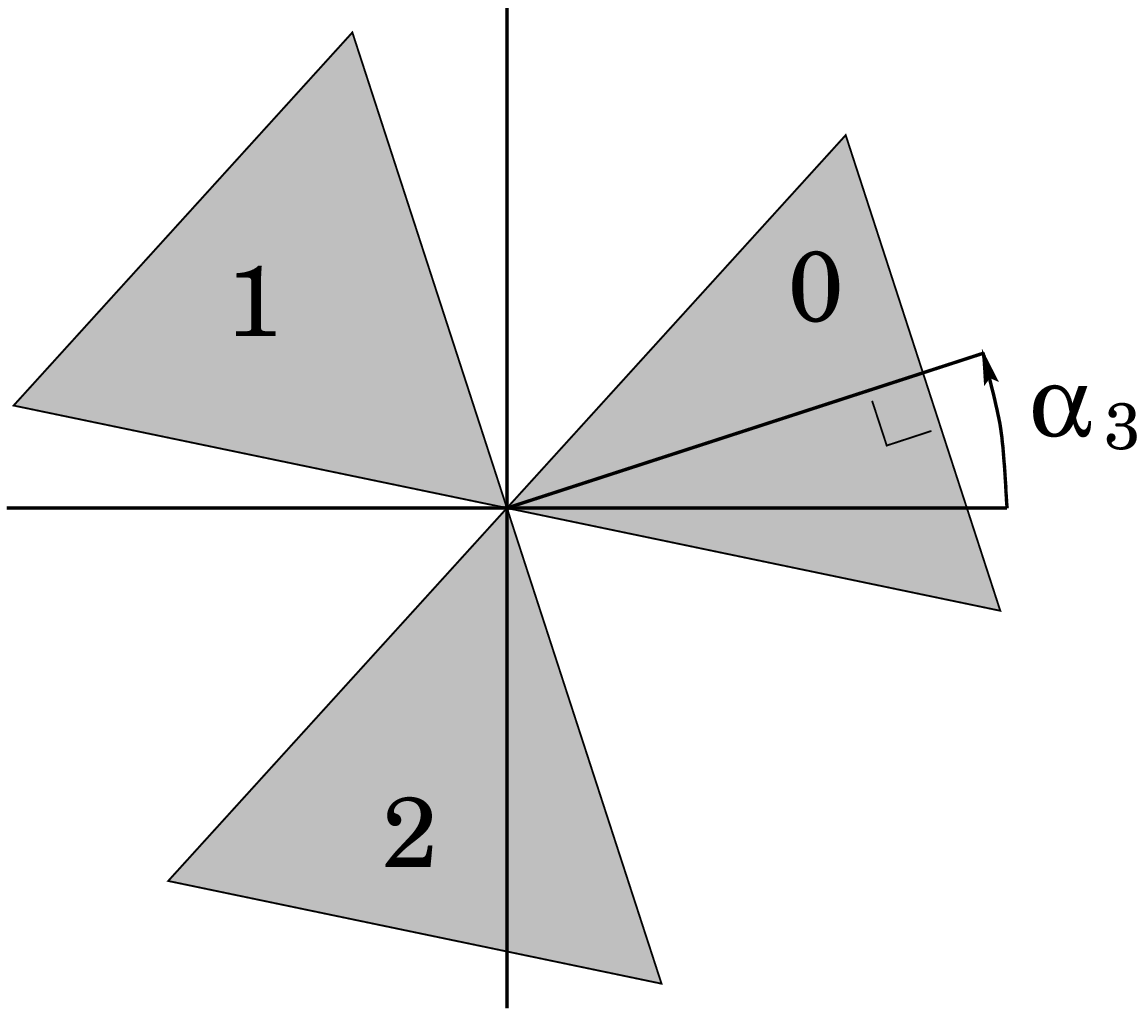,width=0.31\linewidth}
\epsfig{figure=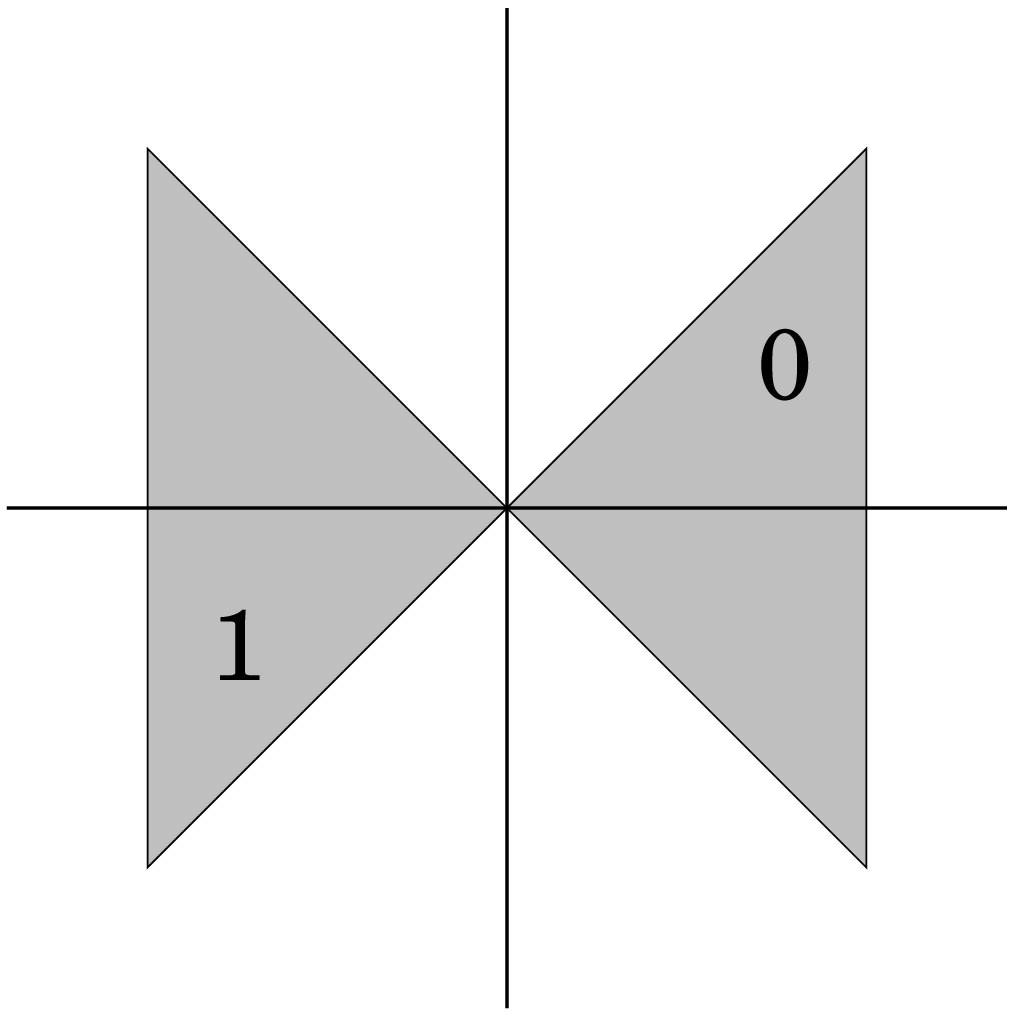,width=0.27\linewidth}
\caption{The regions in the complex $\vhi$-plane which correspond with
         $\Omega_4(k)$, $\Omega_3(k)$ and $\Omega_2(k)$ with 
	 $\eta_j=-j\alpha_j$. }
\label{figcontours}	 
\end{center}
\end{figure}
To illustrate this, let us consider a pure $\vhi^4$-theory, {\it i.e.} with
$\lat=0$. Let the contour start at some $\vhi_1$, chosen at infinity with
argument $\omega_1$, and end at some $\vhi_2$, also at infinity in some
direction with argument $\omega_2$. These values each have to be in some
interval $\Omega_4$: let $\omega_1$ be in $\Omega_4(n_1)$, and $\omega_2$ in
$\Omega(n_2)$. We can sufficiently specify the contour by giving $n_1$ and
$n_2$ so that for instance the contour $\Gamma_{20}$ for $\eta_4=0$ denotes the
standard $\vhi^4$-theory, where we may take the real line for $\Gamma$, start
at $\vhi=-\infty$ and end at $\vhi=+\infty$ (interchange of the endpoints
corresponds to replacing $R$ by $-R$ and hence does not influence $\phi(x)$).
In total, there are six contours that give a viable $\vhi^4$-theory:
$\Gamma_{01}$, $\Gamma_{12}$, $\Gamma_{23}$, $\Gamma_{30}$, $\Gamma_{02}$ and
$\Gamma_{13}$. Note that these are related to each other by phase shifts: in
fact,
\beeqa
\Gamma_{30}=\Gamma_{12}(\eta_4\to\eta_4+2\pi) &,&
\Gamma_{23}=\Gamma_{12}(\eta_4\to\eta_4+4\pi)\;\;,\nl
\Gamma_{12}=\Gamma_{12}(\eta_4\to\eta_4+6\pi) &,&
\Gamma_{13}=\Gamma_{02}(\eta_4\to\eta_4+2\pi)\;\;.
\nn\eneqa
Therefore, only $\Gamma_{02}$ and $\Gamma_{01}$, say, give really different theories,
all other cases being obtainable by an appropriate shift in $\eta_4$.
All contours, as stated, corresponds to viable theories as long as $\laf$ is
non-vanishing, but when we let $|\laf|\to0$ there are two possibilities.
It may happen that $\Omega_4(n_1)$ overlaps with one of the $\Omega_2$
segments, and $\Omega_4(n_2)$ with the {\em other\/} $\Omega_2$ segment. 
In that
case, the limiting theory is equal to the free theory, and the 
limit $|\laf|\to0$ is
smooth: we may call this the perturbative limit. In the other case 
the limit is not smooth, and the path integral $R$ will diverge as 
$|\laf|\to0$:
we call this the non-perturbative limit. Clearly, the limiting 
behavior depends
on the argument $\eta_4$: for the contour $\Gamma_{02}$ 
(the `standard one') one has
\beeqa
-\frac{3}{2}\pi < \eta_4 < \frac{3}{2}\pi\;\;,\;\;
\frac{5}{2}\pi < \eta_4 < \frac{11}{2}\pi &:& \mbox{perturbative}\;\;,\nl
\frac{3}{2}\pi < \eta_4 < \frac{5}{2}\pi\;\;,\;\;
\frac{11}{2}\pi < \eta_4 < \frac{13}{2}\pi &:& \mbox{non-perturbative}\;\;,
\nn\eneqa
and for the other contour $\Gamma_{01}$:
\beeqa
\frac{1}{2}\pi < \eta_4 < \frac{3}{2}\pi\;\;,\;\;
\frac{9}{2}\pi < \eta_4 < \frac{11}{2}\pi &:& \mbox{perturbative}\;\;,\nl
\frac{3}{2}\pi < \eta_4 < \frac{9}{2}\pi\;\;,\;\;
\frac{11}{2}\pi < \eta_4 < \frac{17}{2}\pi &:& \mbox{non-perturbative}\;\;,
\nn\eneqa
For the pure $\vhi^3$-theory, there are of course three contours, related to
each other: $\Gamma_{20}=\Gamma_{01}(\eta_3\to\eta_3+2\pi)$, 
$\Gamma_{12}=\Gamma_{01}(\eta_3\to\eta_3+4\pi)$. For the limiting theory
we find, for contour $\Gamma_{01}$:
\beeqa
-\frac{5}{4}\pi < \eta_3 < \frac{1}{4}\pi\;\;,\;\;
\frac{7}{4}\pi < \eta_3 < \frac{13}{4}\pi &:& \mbox{perturbative}\;\;,\nl
\frac{1}{4}\pi < \eta_3 < \frac{7}{4}\pi\;\;,\;\;
\frac{13}{4}\pi <\eta_3 < \frac{19}{4}\pi &:& \mbox{non-perturbative}\;\;.
\nn\eneqa
In a theory with both $\vhi^3$ and $\vhi^4$ couplings, things become more
interesting. Of course, as long as $\laf$ is nonzero, we are allowed to let
$\lat$ go to zero without jeopardizing the perturbativity. On the other hand,
we can only let $\laf$ vanish with fixed $\lat$ if the selected $\Omega_4$ and
$\Omega_3$ intervals overlap. We give in \fig{shadplot1} and \fig{shadplot2}  
the values of $\eta_4$ and
$\eta_3$ that correspond to a perturbative $\laf\to0$ limit, 
Finally, we may study the combined limits $\laf\to0$ followed by $\lat\to0$. 
The regions of perturbativity are given in \fig{figcontours}. 
Clearly, these are more
restricted since we are in this case requiring a common overlap of 
$\Omega_2$, $\Omega_3$ and $\Omega_4$.

\newcommand{\z}{z}
\newcommand{\bq}{\begin{equation}}
\newcommand{\eq}{  \end{equation}}
\section{Renormalization}

Even though zero-dimensional field theories have no infinities, we may still
consider the effects of renormalization, which here take a graph-theoretical
significance. Renormalizing the field so that the exact propagator is $1/\mm$, 
and the coupling constant so that the proper vertices assume their tree-order
form,
we are counting 
Green's functions without self-energy and vertex insertions, that is, we are
counting the skeleton diagrams of the theory. 
We shall restrict ourselves to theories that are known to be 
perturbatively renormalizable in the usual four-dimensional case.

%
%
\renewcommand{\gt}{g}
\subsection[Renormalization of pure $\vhi^3$-theory]%
           {Renormalization of pure $\boldsymbol{\vhi^3}$-theory}

In this case renormalization proceeds as usual with the introduction of the
tadpole ($\z_1$), the mass ($\z_2$) and the vertex ($\z_3$) counter terms, as
well as the corresponding renormalization conditions that imply the dependence
of these counter terms on the renormalized coupling constant. 
The renormalized action can be written as ($\mu=1,\hbar=1$)
\begin{equation}
  S=\frac{1}{2} \z_2 \vhi^2 +\frac{1}{6} \gt \z_3 \vhi^3 + \z_1 \vhi\;\;,
\end{equation}
and the \SD\ equation takes the form
\begin{equation}
\z_2\phi=(x-\z_1)-\frac{G_3}{2}(\phi^2+\phi') \quad,\qquad G_3\equiv\gt\z_3\;\;.
\end{equation}
Moreover using \eqn{newsteppingeq} we get,
\begin{equation}
  3G_3\frac{\partial\phi}{\partial G_3}=
  \z_2\phi''+2\z_2\phi\phi'-\phi-(x-z_1)\phi' \;\;,
\end{equation}
whereas \eqn{s2} becomes,
\begin{equation}
  \frac{\partial\phi}{\partial \z_2}=-\phi\phi'-\frac{1}{2}\phi'' \;\;.
\label{s2z2}  
\end{equation}
The renormalization conditions that have to be applied are

\BeginCon
\Condition\label{Con31}
No tadpoles, {\it i.e.} $\phi(x=0)=0$ ;

\Condition\label{Con32}
propagator $=\phi'(0)=1$ ;

\Condition\label{Con33}
vertex $=\phi''(0)=-\gt$.
\EndCon

\noindent
Application of these conditions to the \SD\ equation and its derivative leads 
to the equations
\begin{equation}
   \z_1 = -\half\gt\z_3 \quad,\qquad
   \z_2 = 1 + \half\gt^2\z_3 \;\;.
\label{ReEq333}   
\end{equation}
So if we know $\z_3$ as function of $\gt$, we know $\z_1$ and $\z_2$ as
function of $\gt$, and we can consider $\phi$ to be a function of $\gt$ and $x$
only. Its derivative w.r.t.~$\gt$ can, using $\pa/\pa \z_1=-\pa/\pa x$, be
written as
\begin{equation}
\frac{\partial\phi}{\partial\gt}=-\phi'\dot{\z_1}+
\frac{\partial\phi}{\partial \z_2} \dot{\z_2}+
\frac{\partial\phi}{\partial G_3} \dot{G_3}
\end{equation}
where a dot denotes differentiation w.r.t.~$\gt$. Because $\phi$ is a function
of $x$ and $\gt$ only, the l.h.s.~is zero in $x=0$ by \Con{Con31}, and
evaluation of the r.h.s.~leads to the equation
\begin{equation}
   \frac{\gt}{2}\frac{d\z_3}{d\gt}
      = \frac{2\z_3-2(1+\gt^2)\z_3^2}{-4+(4+\gt^2)\z_3} \;\;.
\label{ReEq005}	      
\end{equation}
It is straightforward to derive that for $\gt=0$ the 
perturbative counter terms read
\[
\z_1(0)=0\;,\;\;\z_2(0)=1\;,\;\;\z_3(0)=1\;.
\]
\eqn{ReEq005} is an Abel equation of the second 
kind~\cite{Zwillinger}. The perturbative 
solution, 
satisfying the above initial condition, is 
\begin{equation}
   \z_3(\gt) = 1 - \gt^2 - \frac{1}{2}\gt^4 - 4\gt^6 - 29\gt^8 
                 - \frac{545}{2}\gt^{10} + \cdots\;\;,
\end{equation}
an expansion previously given by Cvitanovi\'c {\it et al.} \cite{Cvitanovic}.

Having, however, solved the \SD\ equation for $\vhi^3$-theory (\Sec{Sec3}), we 
also have an exact, albeit implicit, solution of \eqn{ReEq005},
making use of \Con{Con32}:
\begin{equation}
   [c_1\Ai'(t_0)+c_2\Bi'(t_0)]
\left(\frac{2}{\gt\z_3}\right)^{\!1/3}
   -\frac{\z_2}{\gt\z_3}[c_1\Ai(t_0)+c_2\Bi(t_0)] = 0\;\;,
\label{ReEq008}   
\end{equation}
where
\begin{equation}
   t_0
       = \left(\frac{2}{\gt\z_3}\right)^{\!1/3}
         \left(\half\gt\z_3+\frac{\z_2^2}{2\gt\z_3}\right)
       = \frac{\z_2^2}{(2\gt^2\z_3^2)^{2/3}}
         \left(1+\frac{\gt^2\z_3^2}{\z_2^2}\right)  \;\;,
\nn\end{equation}
and $\Ai$ and $\Bi$ are the two independent solutions of the Airy equation
$f''(t)=tf(t)$. The meaning of this equation
\eqn{ReEq008} is that for a given $g$ and by using 
\eqn{ReEq333} as well as the functional form of $\Ai$ and 
$\Bi$ we can determine $\z_3$.
To show that \eqn{ReEq008} is an implicit solution of \eqn{ReEq005}, 
let
\begin{equation}
   F(\gt) = \frac{(2\gt^2\z_3^2)^{1/3}}{\z_2} \;\;,
\nn\end{equation}
implying $t_0F(\gt)^2-1=(\gt\z_3/\z_2)^2$, and differentiate \eqn{ReEq008} with
respect to $\gt$ to get 
\begin{equation}
     \left(F'(\gt)-\frac{dt_0}{d\gt}\right)[c_1\Ai'(t_0)+c_2\Bi'(t_0)]
   + F(\gt)[c_1\Ai''(t_0)+c_2\Bi''(t_0)]\frac{dt_0}{d\gt} = 0\;\;.
\nn\end{equation}
Using the Airy equation and \eqn{ReEq008}, we get
\begin{equation}
   F'(\gt) + \frac{dt_0}{d\gt}(t_0F(\gt)^2-1) = 0 \;\;.
\nn\end{equation}
Explicitly, this says
\begin{equation}
   2\z_2\frac{d(\gt\z_3)}{d\gt} - 3\gt\z_3\frac{d\z_2}{d\gt} 
                                - 2\z_3\frac{d(\gt\z_3)}{d\gt} = 0 \;\;.
\nn\end{equation}
By using (\ref{ReEq333}), one easily sees that the above equation is an
equivalent form of \eqn{ReEq005}.

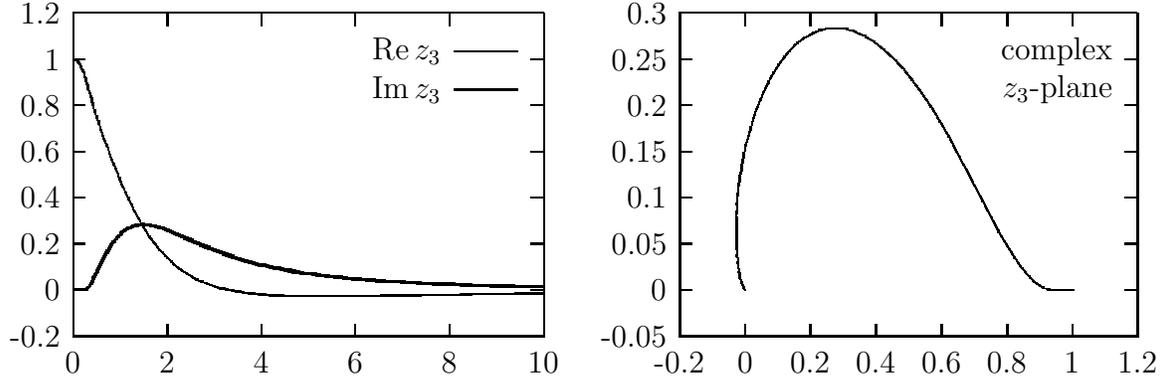
\begin{figure}
\begin{center}
\setlength{\unitlength}{0.240900pt}
\ifx\plotpoint\undefined\newsavebox{\plotpoint}\fi
\sbox{\plotpoint}{\rule[-0.200pt]{0.400pt}{0.400pt}}%
\begin{picture}(900,630)(0,0)
\sbox{\plotpoint}{\rule[-0.200pt]{0.400pt}{0.400pt}}%
\put(140.0,82.0){\rule[-0.200pt]{4.818pt}{0.400pt}}
\put(120,82){\makebox(0,0)[r]{-0.2}}
\put(859.0,82.0){\rule[-0.200pt]{4.818pt}{0.400pt}}
\put(140.0,155.0){\rule[-0.200pt]{4.818pt}{0.400pt}}
\put(120,155){\makebox(0,0)[r]{0}}
\put(859.0,155.0){\rule[-0.200pt]{4.818pt}{0.400pt}}
\put(140.0,227.0){\rule[-0.200pt]{4.818pt}{0.400pt}}
\put(120,227){\makebox(0,0)[r]{0.2}}
\put(859.0,227.0){\rule[-0.200pt]{4.818pt}{0.400pt}}
\put(140.0,300.0){\rule[-0.200pt]{4.818pt}{0.400pt}}
\put(120,300){\makebox(0,0)[r]{0.4}}
\put(859.0,300.0){\rule[-0.200pt]{4.818pt}{0.400pt}}
\put(140.0,372.0){\rule[-0.200pt]{4.818pt}{0.400pt}}
\put(120,372){\makebox(0,0)[r]{0.6}}
\put(859.0,372.0){\rule[-0.200pt]{4.818pt}{0.400pt}}
\put(140.0,445.0){\rule[-0.200pt]{4.818pt}{0.400pt}}
\put(120,445){\makebox(0,0)[r]{0.8}}
\put(859.0,445.0){\rule[-0.200pt]{4.818pt}{0.400pt}}
\put(140.0,517.0){\rule[-0.200pt]{4.818pt}{0.400pt}}
\put(120,517){\makebox(0,0)[r]{1}}
\put(859.0,517.0){\rule[-0.200pt]{4.818pt}{0.400pt}}
\put(140.0,590.0){\rule[-0.200pt]{4.818pt}{0.400pt}}
\put(120,590){\makebox(0,0)[r]{1.2}}
\put(859.0,590.0){\rule[-0.200pt]{4.818pt}{0.400pt}}
\put(140.0,82.0){\rule[-0.200pt]{0.400pt}{4.818pt}}
\put(140,41){\makebox(0,0){0}}
\put(140.0,570.0){\rule[-0.200pt]{0.400pt}{4.818pt}}
\put(288.0,82.0){\rule[-0.200pt]{0.400pt}{4.818pt}}
\put(288,41){\makebox(0,0){2}}
\put(288.0,570.0){\rule[-0.200pt]{0.400pt}{4.818pt}}
\put(436.0,82.0){\rule[-0.200pt]{0.400pt}{4.818pt}}
\put(436,41){\makebox(0,0){4}}
\put(436.0,570.0){\rule[-0.200pt]{0.400pt}{4.818pt}}
\put(583.0,82.0){\rule[-0.200pt]{0.400pt}{4.818pt}}
\put(583,41){\makebox(0,0){6}}
\put(583.0,570.0){\rule[-0.200pt]{0.400pt}{4.818pt}}
\put(731.0,82.0){\rule[-0.200pt]{0.400pt}{4.818pt}}
\put(731,41){\makebox(0,0){8}}
\put(731.0,570.0){\rule[-0.200pt]{0.400pt}{4.818pt}}
\put(879.0,82.0){\rule[-0.200pt]{0.400pt}{4.818pt}}
\put(879,41){\makebox(0,0){10}}
\put(879.0,570.0){\rule[-0.200pt]{0.400pt}{4.818pt}}
\put(140.0,82.0){\rule[-0.200pt]{178.025pt}{0.400pt}}
\put(879.0,82.0){\rule[-0.200pt]{0.400pt}{122.377pt}}
\put(140.0,590.0){\rule[-0.200pt]{178.025pt}{0.400pt}}
\put(140.0,82.0){\rule[-0.200pt]{0.400pt}{122.377pt}}
\put(719,530){\makebox(0,0)[r]{$\mbox{Re}\,\z_3$}}
\put(739.0,530.0){\rule[-0.200pt]{24.090pt}{0.400pt}}
\put(140,518){\usebox{\plotpoint}}
\put(140,518){\usebox{\plotpoint}}
\put(140,518){\usebox{\plotpoint}}
\put(140.0,517.0){\usebox{\plotpoint}}
\put(143,515.67){\rule{0.241pt}{0.400pt}}
\multiput(143.00,516.17)(0.500,-1.000){2}{\rule{0.120pt}{0.400pt}}
\put(140.0,517.0){\rule[-0.200pt]{0.723pt}{0.400pt}}
\put(144,516){\usebox{\plotpoint}}
\put(144,516){\usebox{\plotpoint}}
\put(144,516){\usebox{\plotpoint}}
\put(144,516){\usebox{\plotpoint}}
\put(144,516){\usebox{\plotpoint}}
\put(144.0,516.0){\usebox{\plotpoint}}
\put(145.0,515.0){\usebox{\plotpoint}}
\put(145.0,515.0){\rule[-0.200pt]{0.482pt}{0.400pt}}
\put(147,512.67){\rule{0.241pt}{0.400pt}}
\multiput(147.00,513.17)(0.500,-1.000){2}{\rule{0.120pt}{0.400pt}}
\put(147.0,514.0){\usebox{\plotpoint}}
\put(148,513){\usebox{\plotpoint}}
\put(148,513){\usebox{\plotpoint}}
\put(148,513){\usebox{\plotpoint}}
\put(148,511.67){\rule{0.241pt}{0.400pt}}
\multiput(148.00,512.17)(0.500,-1.000){2}{\rule{0.120pt}{0.400pt}}
\put(149,512){\usebox{\plotpoint}}
\put(149,512){\usebox{\plotpoint}}
\put(148.67,510){\rule{0.400pt}{0.482pt}}
\multiput(148.17,511.00)(1.000,-1.000){2}{\rule{0.400pt}{0.241pt}}
\put(150,508.17){\rule{0.482pt}{0.400pt}}
\multiput(150.00,509.17)(1.000,-2.000){2}{\rule{0.241pt}{0.400pt}}
\put(151.67,506){\rule{0.400pt}{0.482pt}}
\multiput(151.17,507.00)(1.000,-1.000){2}{\rule{0.400pt}{0.241pt}}
\put(152.67,503){\rule{0.400pt}{0.723pt}}
\multiput(152.17,504.50)(1.000,-1.500){2}{\rule{0.400pt}{0.361pt}}
\put(154.17,500){\rule{0.400pt}{0.700pt}}
\multiput(153.17,501.55)(2.000,-1.547){2}{\rule{0.400pt}{0.350pt}}
\put(156.17,496){\rule{0.400pt}{0.900pt}}
\multiput(155.17,498.13)(2.000,-2.132){2}{\rule{0.400pt}{0.450pt}}
\put(157.67,491){\rule{0.400pt}{1.204pt}}
\multiput(157.17,493.50)(1.000,-2.500){2}{\rule{0.400pt}{0.602pt}}
\put(159.17,485){\rule{0.400pt}{1.300pt}}
\multiput(158.17,488.30)(2.000,-3.302){2}{\rule{0.400pt}{0.650pt}}
\put(161.17,478){\rule{0.400pt}{1.500pt}}
\multiput(160.17,481.89)(2.000,-3.887){2}{\rule{0.400pt}{0.750pt}}
\put(163.17,470){\rule{0.400pt}{1.700pt}}
\multiput(162.17,474.47)(2.000,-4.472){2}{\rule{0.400pt}{0.850pt}}
\multiput(165.61,465.16)(0.447,-1.579){3}{\rule{0.108pt}{1.167pt}}
\multiput(164.17,467.58)(3.000,-5.579){2}{\rule{0.400pt}{0.583pt}}
\put(168.17,453){\rule{0.400pt}{1.900pt}}
\multiput(167.17,458.06)(2.000,-5.056){2}{\rule{0.400pt}{0.950pt}}
\put(170.17,445){\rule{0.400pt}{1.700pt}}
\multiput(169.17,449.47)(2.000,-4.472){2}{\rule{0.400pt}{0.850pt}}
\multiput(172.61,439.60)(0.447,-1.802){3}{\rule{0.108pt}{1.300pt}}
\multiput(171.17,442.30)(3.000,-6.302){2}{\rule{0.400pt}{0.650pt}}
\multiput(175.61,430.60)(0.447,-1.802){3}{\rule{0.108pt}{1.300pt}}
\multiput(174.17,433.30)(3.000,-6.302){2}{\rule{0.400pt}{0.650pt}}
\multiput(178.61,421.60)(0.447,-1.802){3}{\rule{0.108pt}{1.300pt}}
\multiput(177.17,424.30)(3.000,-6.302){2}{\rule{0.400pt}{0.650pt}}
\multiput(181.61,412.60)(0.447,-1.802){3}{\rule{0.108pt}{1.300pt}}
\multiput(180.17,415.30)(3.000,-6.302){2}{\rule{0.400pt}{0.650pt}}
\multiput(184.61,403.60)(0.447,-1.802){3}{\rule{0.108pt}{1.300pt}}
\multiput(183.17,406.30)(3.000,-6.302){2}{\rule{0.400pt}{0.650pt}}
\multiput(187.61,394.05)(0.447,-2.025){3}{\rule{0.108pt}{1.433pt}}
\multiput(186.17,397.03)(3.000,-7.025){2}{\rule{0.400pt}{0.717pt}}
\multiput(190.60,385.85)(0.468,-1.212){5}{\rule{0.113pt}{1.000pt}}
\multiput(189.17,387.92)(4.000,-6.924){2}{\rule{0.400pt}{0.500pt}}
\multiput(194.60,376.43)(0.468,-1.358){5}{\rule{0.113pt}{1.100pt}}
\multiput(193.17,378.72)(4.000,-7.717){2}{\rule{0.400pt}{0.550pt}}
\multiput(198.61,365.05)(0.447,-2.025){3}{\rule{0.108pt}{1.433pt}}
\multiput(197.17,368.03)(3.000,-7.025){2}{\rule{0.400pt}{0.717pt}}
\multiput(201.60,356.43)(0.468,-1.358){5}{\rule{0.113pt}{1.100pt}}
\multiput(200.17,358.72)(4.000,-7.717){2}{\rule{0.400pt}{0.550pt}}
\multiput(205.59,346.93)(0.477,-1.155){7}{\rule{0.115pt}{0.980pt}}
\multiput(204.17,348.97)(5.000,-8.966){2}{\rule{0.400pt}{0.490pt}}
\multiput(210.60,335.43)(0.468,-1.358){5}{\rule{0.113pt}{1.100pt}}
\multiput(209.17,337.72)(4.000,-7.717){2}{\rule{0.400pt}{0.550pt}}
\multiput(214.60,325.02)(0.468,-1.505){5}{\rule{0.113pt}{1.200pt}}
\multiput(213.17,327.51)(4.000,-8.509){2}{\rule{0.400pt}{0.600pt}}
\multiput(218.59,315.26)(0.477,-1.044){7}{\rule{0.115pt}{0.900pt}}
\multiput(217.17,317.13)(5.000,-8.132){2}{\rule{0.400pt}{0.450pt}}
\multiput(223.59,304.93)(0.477,-1.155){7}{\rule{0.115pt}{0.980pt}}
\multiput(222.17,306.97)(5.000,-8.966){2}{\rule{0.400pt}{0.490pt}}
\multiput(228.59,294.26)(0.477,-1.044){7}{\rule{0.115pt}{0.900pt}}
\multiput(227.17,296.13)(5.000,-8.132){2}{\rule{0.400pt}{0.450pt}}
\multiput(233.59,283.93)(0.477,-1.155){7}{\rule{0.115pt}{0.980pt}}
\multiput(232.17,285.97)(5.000,-8.966){2}{\rule{0.400pt}{0.490pt}}
\multiput(238.59,273.82)(0.482,-0.852){9}{\rule{0.116pt}{0.767pt}}
\multiput(237.17,275.41)(6.000,-8.409){2}{\rule{0.400pt}{0.383pt}}
\multiput(244.59,263.26)(0.477,-1.044){7}{\rule{0.115pt}{0.900pt}}
\multiput(243.17,265.13)(5.000,-8.132){2}{\rule{0.400pt}{0.450pt}}
\multiput(249.59,254.09)(0.482,-0.762){9}{\rule{0.116pt}{0.700pt}}
\multiput(248.17,255.55)(6.000,-7.547){2}{\rule{0.400pt}{0.350pt}}
\multiput(255.59,244.82)(0.482,-0.852){9}{\rule{0.116pt}{0.767pt}}
\multiput(254.17,246.41)(6.000,-8.409){2}{\rule{0.400pt}{0.383pt}}
\multiput(261.59,235.45)(0.485,-0.645){11}{\rule{0.117pt}{0.614pt}}
\multiput(260.17,236.73)(7.000,-7.725){2}{\rule{0.400pt}{0.307pt}}
\multiput(268.59,226.37)(0.482,-0.671){9}{\rule{0.116pt}{0.633pt}}
\multiput(267.17,227.69)(6.000,-6.685){2}{\rule{0.400pt}{0.317pt}}
\multiput(274.59,218.69)(0.485,-0.569){11}{\rule{0.117pt}{0.557pt}}
\multiput(273.17,219.84)(7.000,-6.844){2}{\rule{0.400pt}{0.279pt}}
\multiput(281.00,211.93)(0.492,-0.485){11}{\rule{0.500pt}{0.117pt}}
\multiput(281.00,212.17)(5.962,-7.000){2}{\rule{0.250pt}{0.400pt}}
\multiput(288.00,204.93)(0.492,-0.485){11}{\rule{0.500pt}{0.117pt}}
\multiput(288.00,205.17)(5.962,-7.000){2}{\rule{0.250pt}{0.400pt}}
\multiput(295.00,197.93)(0.492,-0.485){11}{\rule{0.500pt}{0.117pt}}
\multiput(295.00,198.17)(5.962,-7.000){2}{\rule{0.250pt}{0.400pt}}
\multiput(302.00,190.93)(0.671,-0.482){9}{\rule{0.633pt}{0.116pt}}
\multiput(302.00,191.17)(6.685,-6.000){2}{\rule{0.317pt}{0.400pt}}
\multiput(310.00,184.93)(0.821,-0.477){7}{\rule{0.740pt}{0.115pt}}
\multiput(310.00,185.17)(6.464,-5.000){2}{\rule{0.370pt}{0.400pt}}
\multiput(318.00,179.93)(0.821,-0.477){7}{\rule{0.740pt}{0.115pt}}
\multiput(318.00,180.17)(6.464,-5.000){2}{\rule{0.370pt}{0.400pt}}
\multiput(326.00,174.93)(0.821,-0.477){7}{\rule{0.740pt}{0.115pt}}
\multiput(326.00,175.17)(6.464,-5.000){2}{\rule{0.370pt}{0.400pt}}
\multiput(334.00,169.95)(1.802,-0.447){3}{\rule{1.300pt}{0.108pt}}
\multiput(334.00,170.17)(6.302,-3.000){2}{\rule{0.650pt}{0.400pt}}
\multiput(343.00,166.94)(1.212,-0.468){5}{\rule{1.000pt}{0.113pt}}
\multiput(343.00,167.17)(6.924,-4.000){2}{\rule{0.500pt}{0.400pt}}
\multiput(352.00,162.95)(1.802,-0.447){3}{\rule{1.300pt}{0.108pt}}
\multiput(352.00,163.17)(6.302,-3.000){2}{\rule{0.650pt}{0.400pt}}
\multiput(361.00,159.95)(1.802,-0.447){3}{\rule{1.300pt}{0.108pt}}
\multiput(361.00,160.17)(6.302,-3.000){2}{\rule{0.650pt}{0.400pt}}
\put(370,156.17){\rule{2.100pt}{0.400pt}}
\multiput(370.00,157.17)(5.641,-2.000){2}{\rule{1.050pt}{0.400pt}}
\put(380,154.17){\rule{1.900pt}{0.400pt}}
\multiput(380.00,155.17)(5.056,-2.000){2}{\rule{0.950pt}{0.400pt}}
\put(389,152.17){\rule{2.300pt}{0.400pt}}
\multiput(389.00,153.17)(6.226,-2.000){2}{\rule{1.150pt}{0.400pt}}
\put(400,150.17){\rule{2.100pt}{0.400pt}}
\multiput(400.00,151.17)(5.641,-2.000){2}{\rule{1.050pt}{0.400pt}}
\put(410,148.67){\rule{2.650pt}{0.400pt}}
\multiput(410.00,149.17)(5.500,-1.000){2}{\rule{1.325pt}{0.400pt}}
\put(421,147.67){\rule{2.409pt}{0.400pt}}
\multiput(421.00,148.17)(5.000,-1.000){2}{\rule{1.204pt}{0.400pt}}
\put(431,146.67){\rule{2.891pt}{0.400pt}}
\multiput(431.00,147.17)(6.000,-1.000){2}{\rule{1.445pt}{0.400pt}}
\put(443,145.67){\rule{2.650pt}{0.400pt}}
\multiput(443.00,146.17)(5.500,-1.000){2}{\rule{1.325pt}{0.400pt}}
\put(466,144.67){\rule{2.891pt}{0.400pt}}
\multiput(466.00,145.17)(6.000,-1.000){2}{\rule{1.445pt}{0.400pt}}
\put(454.0,146.0){\rule[-0.200pt]{2.891pt}{0.400pt}}
\put(616,144.67){\rule{3.614pt}{0.400pt}}
\multiput(616.00,144.17)(7.500,1.000){2}{\rule{1.807pt}{0.400pt}}
\put(478.0,145.0){\rule[-0.200pt]{33.244pt}{0.400pt}}
\put(680,145.67){\rule{3.854pt}{0.400pt}}
\multiput(680.00,145.17)(8.000,1.000){2}{\rule{1.927pt}{0.400pt}}
\put(631.0,146.0){\rule[-0.200pt]{11.804pt}{0.400pt}}
\put(749,146.67){\rule{4.336pt}{0.400pt}}
\multiput(749.00,146.17)(9.000,1.000){2}{\rule{2.168pt}{0.400pt}}
\put(696.0,147.0){\rule[-0.200pt]{12.768pt}{0.400pt}}
\put(824,147.67){\rule{4.818pt}{0.400pt}}
\multiput(824.00,147.17)(10.000,1.000){2}{\rule{2.409pt}{0.400pt}}
\put(767.0,148.0){\rule[-0.200pt]{13.731pt}{0.400pt}}
\put(844.0,149.0){\rule[-0.200pt]{8.431pt}{0.400pt}}
\sbox{\plotpoint}{\rule[-0.400pt]{0.800pt}{0.800pt}}%
\put(719,470){\makebox(0,0)[r]{$\mbox{Im}\,\z_3$}}
\put(739.0,470.0){\rule[-0.400pt]{24.090pt}{0.800pt}}
\put(140,155){\usebox{\plotpoint}}
\put(140,155){\usebox{\plotpoint}}
\put(140,155){\usebox{\plotpoint}}
\put(140,155){\usebox{\plotpoint}}
\put(140,155){\usebox{\plotpoint}}
\put(159,153.84){\rule{0.482pt}{0.800pt}}
\multiput(159.00,153.34)(1.000,1.000){2}{\rule{0.241pt}{0.800pt}}
\put(161,154.84){\rule{0.482pt}{0.800pt}}
\multiput(161.00,154.34)(1.000,1.000){2}{\rule{0.241pt}{0.800pt}}
\put(162.34,157){\rule{0.800pt}{0.723pt}}
\multiput(161.34,157.00)(2.000,1.500){2}{\rule{0.800pt}{0.361pt}}
\put(165,159.84){\rule{0.723pt}{0.800pt}}
\multiput(165.00,158.34)(1.500,3.000){2}{\rule{0.361pt}{0.800pt}}
\put(167.34,163){\rule{0.800pt}{0.964pt}}
\multiput(166.34,163.00)(2.000,2.000){2}{\rule{0.800pt}{0.482pt}}
\put(169.34,167){\rule{0.800pt}{1.204pt}}
\multiput(168.34,167.00)(2.000,2.500){2}{\rule{0.800pt}{0.602pt}}
\put(171.84,172){\rule{0.800pt}{1.445pt}}
\multiput(170.34,172.00)(3.000,3.000){2}{\rule{0.800pt}{0.723pt}}
\put(174.84,178){\rule{0.800pt}{1.204pt}}
\multiput(173.34,178.00)(3.000,2.500){2}{\rule{0.800pt}{0.602pt}}
\put(177.84,183){\rule{0.800pt}{1.686pt}}
\multiput(176.34,183.00)(3.000,3.500){2}{\rule{0.800pt}{0.843pt}}
\put(180.84,190){\rule{0.800pt}{1.445pt}}
\multiput(179.34,190.00)(3.000,3.000){2}{\rule{0.800pt}{0.723pt}}
\put(183.84,196){\rule{0.800pt}{1.445pt}}
\multiput(182.34,196.00)(3.000,3.000){2}{\rule{0.800pt}{0.723pt}}
\put(186.84,202){\rule{0.800pt}{1.686pt}}
\multiput(185.34,202.00)(3.000,3.500){2}{\rule{0.800pt}{0.843pt}}
\put(190.34,209){\rule{0.800pt}{1.400pt}}
\multiput(188.34,209.00)(4.000,3.094){2}{\rule{0.800pt}{0.700pt}}
\put(194.34,215){\rule{0.800pt}{1.400pt}}
\multiput(192.34,215.00)(4.000,3.094){2}{\rule{0.800pt}{0.700pt}}
\put(197.84,221){\rule{0.800pt}{1.445pt}}
\multiput(196.34,221.00)(3.000,3.000){2}{\rule{0.800pt}{0.723pt}}
\put(201.34,227){\rule{0.800pt}{1.200pt}}
\multiput(199.34,227.00)(4.000,2.509){2}{\rule{0.800pt}{0.600pt}}
\multiput(205.00,233.38)(0.424,0.560){3}{\rule{1.000pt}{0.135pt}}
\multiput(205.00,230.34)(2.924,5.000){2}{\rule{0.500pt}{0.800pt}}
\put(210.34,237){\rule{0.800pt}{1.200pt}}
\multiput(208.34,237.00)(4.000,2.509){2}{\rule{0.800pt}{0.600pt}}
\put(214,242.34){\rule{0.964pt}{0.800pt}}
\multiput(214.00,240.34)(2.000,4.000){2}{\rule{0.482pt}{0.800pt}}
\put(218,245.84){\rule{1.204pt}{0.800pt}}
\multiput(218.00,244.34)(2.500,3.000){2}{\rule{0.602pt}{0.800pt}}
\put(223,248.84){\rule{1.204pt}{0.800pt}}
\multiput(223.00,247.34)(2.500,3.000){2}{\rule{0.602pt}{0.800pt}}
\put(228,251.34){\rule{1.204pt}{0.800pt}}
\multiput(228.00,250.34)(2.500,2.000){2}{\rule{0.602pt}{0.800pt}}
\put(233,253.34){\rule{1.204pt}{0.800pt}}
\multiput(233.00,252.34)(2.500,2.000){2}{\rule{0.602pt}{0.800pt}}
\put(238,254.84){\rule{1.445pt}{0.800pt}}
\multiput(238.00,254.34)(3.000,1.000){2}{\rule{0.723pt}{0.800pt}}
\put(244,255.84){\rule{1.204pt}{0.800pt}}
\multiput(244.00,255.34)(2.500,1.000){2}{\rule{0.602pt}{0.800pt}}
\put(249,255.84){\rule{1.445pt}{0.800pt}}
\multiput(249.00,256.34)(3.000,-1.000){2}{\rule{0.723pt}{0.800pt}}
\put(140.0,155.0){\rule[-0.400pt]{4.577pt}{0.800pt}}
\put(261,254.34){\rule{1.686pt}{0.800pt}}
\multiput(261.00,255.34)(3.500,-2.000){2}{\rule{0.843pt}{0.800pt}}
\put(268,252.84){\rule{1.445pt}{0.800pt}}
\multiput(268.00,253.34)(3.000,-1.000){2}{\rule{0.723pt}{0.800pt}}
\put(274,251.34){\rule{1.686pt}{0.800pt}}
\multiput(274.00,252.34)(3.500,-2.000){2}{\rule{0.843pt}{0.800pt}}
\put(281,248.84){\rule{1.686pt}{0.800pt}}
\multiput(281.00,250.34)(3.500,-3.000){2}{\rule{0.843pt}{0.800pt}}
\put(288,245.84){\rule{1.686pt}{0.800pt}}
\multiput(288.00,247.34)(3.500,-3.000){2}{\rule{0.843pt}{0.800pt}}
\put(295,242.84){\rule{1.686pt}{0.800pt}}
\multiput(295.00,244.34)(3.500,-3.000){2}{\rule{0.843pt}{0.800pt}}
\put(302,239.84){\rule{1.927pt}{0.800pt}}
\multiput(302.00,241.34)(4.000,-3.000){2}{\rule{0.964pt}{0.800pt}}
\put(310,236.34){\rule{1.800pt}{0.800pt}}
\multiput(310.00,238.34)(4.264,-4.000){2}{\rule{0.900pt}{0.800pt}}
\put(318,232.84){\rule{1.927pt}{0.800pt}}
\multiput(318.00,234.34)(4.000,-3.000){2}{\rule{0.964pt}{0.800pt}}
\put(326,229.34){\rule{1.800pt}{0.800pt}}
\multiput(326.00,231.34)(4.264,-4.000){2}{\rule{0.900pt}{0.800pt}}
\put(334,225.34){\rule{2.000pt}{0.800pt}}
\multiput(334.00,227.34)(4.849,-4.000){2}{\rule{1.000pt}{0.800pt}}
\put(343,221.84){\rule{2.168pt}{0.800pt}}
\multiput(343.00,223.34)(4.500,-3.000){2}{\rule{1.084pt}{0.800pt}}
\put(352,218.34){\rule{2.000pt}{0.800pt}}
\multiput(352.00,220.34)(4.849,-4.000){2}{\rule{1.000pt}{0.800pt}}
\put(361,214.84){\rule{2.168pt}{0.800pt}}
\multiput(361.00,216.34)(4.500,-3.000){2}{\rule{1.084pt}{0.800pt}}
\put(370,211.34){\rule{2.200pt}{0.800pt}}
\multiput(370.00,213.34)(5.434,-4.000){2}{\rule{1.100pt}{0.800pt}}
\put(380,207.84){\rule{2.168pt}{0.800pt}}
\multiput(380.00,209.34)(4.500,-3.000){2}{\rule{1.084pt}{0.800pt}}
\put(389,204.34){\rule{2.400pt}{0.800pt}}
\multiput(389.00,206.34)(6.019,-4.000){2}{\rule{1.200pt}{0.800pt}}
\put(400,200.84){\rule{2.409pt}{0.800pt}}
\multiput(400.00,202.34)(5.000,-3.000){2}{\rule{1.204pt}{0.800pt}}
\put(410,197.84){\rule{2.650pt}{0.800pt}}
\multiput(410.00,199.34)(5.500,-3.000){2}{\rule{1.325pt}{0.800pt}}
\put(421,194.84){\rule{2.409pt}{0.800pt}}
\multiput(421.00,196.34)(5.000,-3.000){2}{\rule{1.204pt}{0.800pt}}
\put(431,192.34){\rule{2.891pt}{0.800pt}}
\multiput(431.00,193.34)(6.000,-2.000){2}{\rule{1.445pt}{0.800pt}}
\put(443,189.84){\rule{2.650pt}{0.800pt}}
\multiput(443.00,191.34)(5.500,-3.000){2}{\rule{1.325pt}{0.800pt}}
\put(454,187.34){\rule{2.891pt}{0.800pt}}
\multiput(454.00,188.34)(6.000,-2.000){2}{\rule{1.445pt}{0.800pt}}
\put(466,185.34){\rule{2.891pt}{0.800pt}}
\multiput(466.00,186.34)(6.000,-2.000){2}{\rule{1.445pt}{0.800pt}}
\put(478,182.84){\rule{2.891pt}{0.800pt}}
\multiput(478.00,184.34)(6.000,-3.000){2}{\rule{1.445pt}{0.800pt}}
\put(490,180.34){\rule{3.132pt}{0.800pt}}
\multiput(490.00,181.34)(6.500,-2.000){2}{\rule{1.566pt}{0.800pt}}
\put(503,178.84){\rule{3.132pt}{0.800pt}}
\multiput(503.00,179.34)(6.500,-1.000){2}{\rule{1.566pt}{0.800pt}}
\put(516,177.34){\rule{3.132pt}{0.800pt}}
\multiput(516.00,178.34)(6.500,-2.000){2}{\rule{1.566pt}{0.800pt}}
\put(529,175.34){\rule{3.373pt}{0.800pt}}
\multiput(529.00,176.34)(7.000,-2.000){2}{\rule{1.686pt}{0.800pt}}
\put(543,173.84){\rule{3.373pt}{0.800pt}}
\multiput(543.00,174.34)(7.000,-1.000){2}{\rule{1.686pt}{0.800pt}}
\put(557,172.34){\rule{3.373pt}{0.800pt}}
\multiput(557.00,173.34)(7.000,-2.000){2}{\rule{1.686pt}{0.800pt}}
\put(571,170.84){\rule{3.614pt}{0.800pt}}
\multiput(571.00,171.34)(7.500,-1.000){2}{\rule{1.807pt}{0.800pt}}
\put(586,169.84){\rule{3.373pt}{0.800pt}}
\multiput(586.00,170.34)(7.000,-1.000){2}{\rule{1.686pt}{0.800pt}}
\put(600,168.84){\rule{3.854pt}{0.800pt}}
\multiput(600.00,169.34)(8.000,-1.000){2}{\rule{1.927pt}{0.800pt}}
\put(616,167.34){\rule{3.614pt}{0.800pt}}
\multiput(616.00,168.34)(7.500,-2.000){2}{\rule{1.807pt}{0.800pt}}
\put(255.0,257.0){\rule[-0.400pt]{1.445pt}{0.800pt}}
\put(647,165.84){\rule{3.854pt}{0.800pt}}
\multiput(647.00,166.34)(8.000,-1.000){2}{\rule{1.927pt}{0.800pt}}
\put(663,164.84){\rule{4.095pt}{0.800pt}}
\multiput(663.00,165.34)(8.500,-1.000){2}{\rule{2.048pt}{0.800pt}}
\put(680,163.84){\rule{3.854pt}{0.800pt}}
\multiput(680.00,164.34)(8.000,-1.000){2}{\rule{1.927pt}{0.800pt}}
\put(696,162.84){\rule{4.336pt}{0.800pt}}
\multiput(696.00,163.34)(9.000,-1.000){2}{\rule{2.168pt}{0.800pt}}
\put(631.0,168.0){\rule[-0.400pt]{3.854pt}{0.800pt}}
\put(731,161.84){\rule{4.336pt}{0.800pt}}
\multiput(731.00,162.34)(9.000,-1.000){2}{\rule{2.168pt}{0.800pt}}
\put(749,160.84){\rule{4.336pt}{0.800pt}}
\multiput(749.00,161.34)(9.000,-1.000){2}{\rule{2.168pt}{0.800pt}}
\put(714.0,164.0){\rule[-0.400pt]{4.095pt}{0.800pt}}
\put(786,159.84){\rule{4.577pt}{0.800pt}}
\multiput(786.00,160.34)(9.500,-1.000){2}{\rule{2.289pt}{0.800pt}}
\put(767.0,162.0){\rule[-0.400pt]{4.577pt}{0.800pt}}
\put(824,158.84){\rule{4.818pt}{0.800pt}}
\multiput(824.00,159.34)(10.000,-1.000){2}{\rule{2.409pt}{0.800pt}}
\put(805.0,161.0){\rule[-0.400pt]{4.577pt}{0.800pt}}
\put(844.0,160.0){\rule[-0.400pt]{8.431pt}{0.800pt}}
\end{picture}
%
%
\setlength{\unitlength}{0.240900pt}
\ifx\plotpoint\undefined\newsavebox{\plotpoint}\fi
\begin{picture}(900,630)(0,0)
\sbox{\plotpoint}{\rule[-0.200pt]{0.400pt}{0.400pt}}%
\put(160.0,82.0){\rule[-0.200pt]{4.818pt}{0.400pt}}
\put(140,82){\makebox(0,0)[r]{-0.05}}
\put(859.0,82.0){\rule[-0.200pt]{4.818pt}{0.400pt}}
\put(160.0,155.0){\rule[-0.200pt]{4.818pt}{0.400pt}}
\put(140,155){\makebox(0,0)[r]{0}}
\put(859.0,155.0){\rule[-0.200pt]{4.818pt}{0.400pt}}
\put(160.0,227.0){\rule[-0.200pt]{4.818pt}{0.400pt}}
\put(140,227){\makebox(0,0)[r]{0.05}}
\put(859.0,227.0){\rule[-0.200pt]{4.818pt}{0.400pt}}
\put(160.0,300.0){\rule[-0.200pt]{4.818pt}{0.400pt}}
\put(140,300){\makebox(0,0)[r]{0.1}}
\put(859.0,300.0){\rule[-0.200pt]{4.818pt}{0.400pt}}
\put(160.0,372.0){\rule[-0.200pt]{4.818pt}{0.400pt}}
\put(140,372){\makebox(0,0)[r]{0.15}}
\put(859.0,372.0){\rule[-0.200pt]{4.818pt}{0.400pt}}
\put(160.0,445.0){\rule[-0.200pt]{4.818pt}{0.400pt}}
\put(140,445){\makebox(0,0)[r]{0.2}}
\put(859.0,445.0){\rule[-0.200pt]{4.818pt}{0.400pt}}
\put(160.0,517.0){\rule[-0.200pt]{4.818pt}{0.400pt}}
\put(140,517){\makebox(0,0)[r]{0.25}}
\put(859.0,517.0){\rule[-0.200pt]{4.818pt}{0.400pt}}
\put(160.0,590.0){\rule[-0.200pt]{4.818pt}{0.400pt}}
\put(140,590){\makebox(0,0)[r]{0.3}}
\put(859.0,590.0){\rule[-0.200pt]{4.818pt}{0.400pt}}
\put(160.0,82.0){\rule[-0.200pt]{0.400pt}{4.818pt}}
\put(160,41){\makebox(0,0){-0.2}}
\put(160.0,570.0){\rule[-0.200pt]{0.400pt}{4.818pt}}
\put(263.0,82.0){\rule[-0.200pt]{0.400pt}{4.818pt}}
\put(263,41){\makebox(0,0){0}}
\put(263.0,570.0){\rule[-0.200pt]{0.400pt}{4.818pt}}
\put(365.0,82.0){\rule[-0.200pt]{0.400pt}{4.818pt}}
\put(365,41){\makebox(0,0){0.2}}
\put(365.0,570.0){\rule[-0.200pt]{0.400pt}{4.818pt}}
\put(468.0,82.0){\rule[-0.200pt]{0.400pt}{4.818pt}}
\put(468,41){\makebox(0,0){0.4}}
\put(468.0,570.0){\rule[-0.200pt]{0.400pt}{4.818pt}}
\put(571.0,82.0){\rule[-0.200pt]{0.400pt}{4.818pt}}
\put(571,41){\makebox(0,0){0.6}}
\put(571.0,570.0){\rule[-0.200pt]{0.400pt}{4.818pt}}
\put(674.0,82.0){\rule[-0.200pt]{0.400pt}{4.818pt}}
\put(674,41){\makebox(0,0){0.8}}
\put(674.0,570.0){\rule[-0.200pt]{0.400pt}{4.818pt}}
\put(776.0,82.0){\rule[-0.200pt]{0.400pt}{4.818pt}}
\put(776,41){\makebox(0,0){1}}
\put(776.0,570.0){\rule[-0.200pt]{0.400pt}{4.818pt}}
\put(879.0,82.0){\rule[-0.200pt]{0.400pt}{4.818pt}}
\put(879,41){\makebox(0,0){1.2}}
\put(879.0,570.0){\rule[-0.200pt]{0.400pt}{4.818pt}}
\put(160.0,82.0){\rule[-0.200pt]{173.207pt}{0.400pt}}
\put(879.0,82.0){\rule[-0.200pt]{0.400pt}{122.377pt}}
\put(160.0,590.0){\rule[-0.200pt]{173.207pt}{0.400pt}}
\put(160.0,82.0){\rule[-0.200pt]{0.400pt}{122.377pt}}
\put(840,530){\makebox(0,0)[r]{complex}}
\put(840,470){\makebox(0,0)[r]{$\z_3$-plane}}
\put(778,155){\usebox{\plotpoint}}
\put(778,155){\usebox{\plotpoint}}
\put(746,153.67){\rule{1.204pt}{0.400pt}}
\multiput(748.50,154.17)(-2.500,-1.000){2}{\rule{0.602pt}{0.400pt}}
\put(738,154.17){\rule{1.700pt}{0.400pt}}
\multiput(742.47,153.17)(-4.472,2.000){2}{\rule{0.850pt}{0.400pt}}
\multiput(733.16,156.61)(-1.579,0.447){3}{\rule{1.167pt}{0.108pt}}
\multiput(735.58,155.17)(-5.579,3.000){2}{\rule{0.583pt}{0.400pt}}
\multiput(727.21,159.59)(-0.721,0.485){11}{\rule{0.671pt}{0.117pt}}
\multiput(728.61,158.17)(-8.606,7.000){2}{\rule{0.336pt}{0.400pt}}
\multiput(717.76,166.58)(-0.547,0.491){17}{\rule{0.540pt}{0.118pt}}
\multiput(718.88,165.17)(-9.879,10.000){2}{\rule{0.270pt}{0.400pt}}
\multiput(707.92,176.00)(-0.492,0.590){19}{\rule{0.118pt}{0.573pt}}
\multiput(708.17,176.00)(-11.000,11.811){2}{\rule{0.400pt}{0.286pt}}
\multiput(696.92,189.00)(-0.492,0.712){21}{\rule{0.119pt}{0.667pt}}
\multiput(697.17,189.00)(-12.000,15.616){2}{\rule{0.400pt}{0.333pt}}
\multiput(684.92,206.00)(-0.492,0.798){21}{\rule{0.119pt}{0.733pt}}
\multiput(685.17,206.00)(-12.000,17.478){2}{\rule{0.400pt}{0.367pt}}
\multiput(672.92,225.00)(-0.493,0.853){23}{\rule{0.119pt}{0.777pt}}
\multiput(673.17,225.00)(-13.000,20.387){2}{\rule{0.400pt}{0.388pt}}
\multiput(659.92,247.00)(-0.492,0.970){21}{\rule{0.119pt}{0.867pt}}
\multiput(660.17,247.00)(-12.000,21.201){2}{\rule{0.400pt}{0.433pt}}
\multiput(647.92,270.00)(-0.493,0.972){23}{\rule{0.119pt}{0.869pt}}
\multiput(648.17,270.00)(-13.000,23.196){2}{\rule{0.400pt}{0.435pt}}
\multiput(634.92,295.00)(-0.493,0.972){23}{\rule{0.119pt}{0.869pt}}
\multiput(635.17,295.00)(-13.000,23.196){2}{\rule{0.400pt}{0.435pt}}
\multiput(621.92,320.00)(-0.493,0.972){23}{\rule{0.119pt}{0.869pt}}
\multiput(622.17,320.00)(-13.000,23.196){2}{\rule{0.400pt}{0.435pt}}
\multiput(608.92,345.00)(-0.494,0.938){25}{\rule{0.119pt}{0.843pt}}
\multiput(609.17,345.00)(-14.000,24.251){2}{\rule{0.400pt}{0.421pt}}
\multiput(594.92,371.00)(-0.493,0.972){23}{\rule{0.119pt}{0.869pt}}
\multiput(595.17,371.00)(-13.000,23.196){2}{\rule{0.400pt}{0.435pt}}
\multiput(581.92,396.00)(-0.494,0.864){25}{\rule{0.119pt}{0.786pt}}
\multiput(582.17,396.00)(-14.000,22.369){2}{\rule{0.400pt}{0.393pt}}
\multiput(567.92,420.00)(-0.494,0.827){25}{\rule{0.119pt}{0.757pt}}
\multiput(568.17,420.00)(-14.000,21.429){2}{\rule{0.400pt}{0.379pt}}
\multiput(553.92,443.00)(-0.494,0.737){27}{\rule{0.119pt}{0.687pt}}
\multiput(554.17,443.00)(-15.000,20.575){2}{\rule{0.400pt}{0.343pt}}
\multiput(538.92,465.00)(-0.494,0.668){27}{\rule{0.119pt}{0.633pt}}
\multiput(539.17,465.00)(-15.000,18.685){2}{\rule{0.400pt}{0.317pt}}
\multiput(523.92,485.00)(-0.494,0.644){25}{\rule{0.119pt}{0.614pt}}
\multiput(524.17,485.00)(-14.000,16.725){2}{\rule{0.400pt}{0.307pt}}
\multiput(509.92,503.00)(-0.494,0.531){27}{\rule{0.119pt}{0.527pt}}
\multiput(510.17,503.00)(-15.000,14.907){2}{\rule{0.400pt}{0.263pt}}
\multiput(493.81,519.58)(-0.534,0.494){25}{\rule{0.529pt}{0.119pt}}
\multiput(494.90,518.17)(-13.903,14.000){2}{\rule{0.264pt}{0.400pt}}
\multiput(478.51,533.58)(-0.625,0.492){21}{\rule{0.600pt}{0.119pt}}
\multiput(479.75,532.17)(-13.755,12.000){2}{\rule{0.300pt}{0.400pt}}
\multiput(462.82,545.59)(-0.844,0.489){15}{\rule{0.767pt}{0.118pt}}
\multiput(464.41,544.17)(-13.409,9.000){2}{\rule{0.383pt}{0.400pt}}
\multiput(447.03,554.59)(-1.103,0.485){11}{\rule{0.957pt}{0.117pt}}
\multiput(449.01,553.17)(-13.013,7.000){2}{\rule{0.479pt}{0.400pt}}
\multiput(429.77,561.60)(-1.943,0.468){5}{\rule{1.500pt}{0.113pt}}
\multiput(432.89,560.17)(-10.887,4.000){2}{\rule{0.750pt}{0.400pt}}
\put(408,564.67){\rule{3.373pt}{0.400pt}}
\multiput(415.00,564.17)(-7.000,1.000){2}{\rule{1.686pt}{0.400pt}}
\put(751.0,155.0){\rule[-0.200pt]{6.504pt}{0.400pt}}
\multiput(386.39,564.95)(-2.695,-0.447){3}{\rule{1.833pt}{0.108pt}}
\multiput(390.19,565.17)(-9.195,-3.000){2}{\rule{0.917pt}{0.400pt}}
\multiput(376.60,561.93)(-1.267,-0.477){7}{\rule{1.060pt}{0.115pt}}
\multiput(378.80,562.17)(-9.800,-5.000){2}{\rule{0.530pt}{0.400pt}}
\multiput(365.74,556.93)(-0.874,-0.485){11}{\rule{0.786pt}{0.117pt}}
\multiput(367.37,557.17)(-10.369,-7.000){2}{\rule{0.393pt}{0.400pt}}
\multiput(354.09,549.93)(-0.758,-0.488){13}{\rule{0.700pt}{0.117pt}}
\multiput(355.55,550.17)(-10.547,-8.000){2}{\rule{0.350pt}{0.400pt}}
\multiput(342.92,541.92)(-0.495,-0.491){17}{\rule{0.500pt}{0.118pt}}
\multiput(343.96,542.17)(-8.962,-10.000){2}{\rule{0.250pt}{0.400pt}}
\multiput(333.92,530.59)(-0.491,-0.600){17}{\rule{0.118pt}{0.580pt}}
\multiput(334.17,531.80)(-10.000,-10.796){2}{\rule{0.400pt}{0.290pt}}
\multiput(323.93,518.37)(-0.489,-0.669){15}{\rule{0.118pt}{0.633pt}}
\multiput(324.17,519.69)(-9.000,-10.685){2}{\rule{0.400pt}{0.317pt}}
\multiput(314.93,505.89)(-0.488,-0.824){13}{\rule{0.117pt}{0.750pt}}
\multiput(315.17,507.44)(-8.000,-11.443){2}{\rule{0.400pt}{0.375pt}}
\multiput(306.93,492.68)(-0.488,-0.890){13}{\rule{0.117pt}{0.800pt}}
\multiput(307.17,494.34)(-8.000,-12.340){2}{\rule{0.400pt}{0.400pt}}
\multiput(298.93,478.03)(-0.485,-1.103){11}{\rule{0.117pt}{0.957pt}}
\multiput(299.17,480.01)(-7.000,-13.013){2}{\rule{0.400pt}{0.479pt}}
\multiput(291.93,462.71)(-0.482,-1.214){9}{\rule{0.116pt}{1.033pt}}
\multiput(292.17,464.86)(-6.000,-11.855){2}{\rule{0.400pt}{0.517pt}}
\multiput(285.93,448.43)(-0.482,-1.304){9}{\rule{0.116pt}{1.100pt}}
\multiput(286.17,450.72)(-6.000,-12.717){2}{\rule{0.400pt}{0.550pt}}
\multiput(279.93,432.60)(-0.477,-1.601){7}{\rule{0.115pt}{1.300pt}}
\multiput(280.17,435.30)(-5.000,-12.302){2}{\rule{0.400pt}{0.650pt}}
\multiput(274.94,416.77)(-0.468,-1.943){5}{\rule{0.113pt}{1.500pt}}
\multiput(275.17,419.89)(-4.000,-10.887){2}{\rule{0.400pt}{0.750pt}}
\multiput(270.94,402.36)(-0.468,-2.090){5}{\rule{0.113pt}{1.600pt}}
\multiput(271.17,405.68)(-4.000,-11.679){2}{\rule{0.400pt}{0.800pt}}
\multiput(266.94,388.19)(-0.468,-1.797){5}{\rule{0.113pt}{1.400pt}}
\multiput(267.17,391.09)(-4.000,-10.094){2}{\rule{0.400pt}{0.700pt}}
\multiput(262.95,372.84)(-0.447,-2.918){3}{\rule{0.108pt}{1.967pt}}
\multiput(263.17,376.92)(-3.000,-9.918){2}{\rule{0.400pt}{0.983pt}}
\put(259.17,354){\rule{0.400pt}{2.700pt}}
\multiput(260.17,361.40)(-2.000,-7.396){2}{\rule{0.400pt}{1.350pt}}
\put(257.17,342){\rule{0.400pt}{2.500pt}}
\multiput(258.17,348.81)(-2.000,-6.811){2}{\rule{0.400pt}{1.250pt}}
\put(255.17,330){\rule{0.400pt}{2.500pt}}
\multiput(256.17,336.81)(-2.000,-6.811){2}{\rule{0.400pt}{1.250pt}}
\put(253.17,318){\rule{0.400pt}{2.500pt}}
\multiput(254.17,324.81)(-2.000,-6.811){2}{\rule{0.400pt}{1.250pt}}
\put(251.67,307){\rule{0.400pt}{2.650pt}}
\multiput(252.17,312.50)(-1.000,-5.500){2}{\rule{0.400pt}{1.325pt}}
\put(250.67,297){\rule{0.400pt}{2.409pt}}
\multiput(251.17,302.00)(-1.000,-5.000){2}{\rule{0.400pt}{1.204pt}}
\put(249.67,288){\rule{0.400pt}{2.168pt}}
\multiput(250.17,292.50)(-1.000,-4.500){2}{\rule{0.400pt}{1.084pt}}
\put(394.0,566.0){\rule[-0.200pt]{3.373pt}{0.400pt}}
\put(248.67,270){\rule{0.400pt}{1.927pt}}
\multiput(249.17,274.00)(-1.000,-4.000){2}{\rule{0.400pt}{0.964pt}}
\put(250.0,278.0){\rule[-0.200pt]{0.400pt}{2.409pt}}
\put(248.67,219){\rule{0.400pt}{1.204pt}}
\multiput(248.17,221.50)(1.000,-2.500){2}{\rule{0.400pt}{0.602pt}}
\put(249.0,224.0){\rule[-0.200pt]{0.400pt}{11.081pt}}
\put(249.67,206){\rule{0.400pt}{0.964pt}}
\multiput(249.17,208.00)(1.000,-2.000){2}{\rule{0.400pt}{0.482pt}}
\put(250.0,210.0){\rule[-0.200pt]{0.400pt}{2.168pt}}
\put(250.67,196){\rule{0.400pt}{0.723pt}}
\multiput(250.17,197.50)(1.000,-1.500){2}{\rule{0.400pt}{0.361pt}}
\put(251.0,199.0){\rule[-0.200pt]{0.400pt}{1.686pt}}
\put(251.67,188){\rule{0.400pt}{0.723pt}}
\multiput(251.17,189.50)(1.000,-1.500){2}{\rule{0.400pt}{0.361pt}}
\put(252.0,191.0){\rule[-0.200pt]{0.400pt}{1.204pt}}
\put(252.67,182){\rule{0.400pt}{0.482pt}}
\multiput(252.17,183.00)(1.000,-1.000){2}{\rule{0.400pt}{0.241pt}}
\put(253.0,184.0){\rule[-0.200pt]{0.400pt}{0.964pt}}
\put(253.67,176){\rule{0.400pt}{0.482pt}}
\multiput(253.17,177.00)(1.000,-1.000){2}{\rule{0.400pt}{0.241pt}}
\put(254.0,178.0){\rule[-0.200pt]{0.400pt}{0.964pt}}
\put(254.67,172){\rule{0.400pt}{0.482pt}}
\multiput(254.17,173.00)(1.000,-1.000){2}{\rule{0.400pt}{0.241pt}}
\put(255.0,174.0){\rule[-0.200pt]{0.400pt}{0.482pt}}
\put(256,167.67){\rule{0.241pt}{0.400pt}}
\multiput(256.00,168.17)(0.500,-1.000){2}{\rule{0.120pt}{0.400pt}}
\put(256.0,169.0){\rule[-0.200pt]{0.400pt}{0.723pt}}
\put(257,164.67){\rule{0.241pt}{0.400pt}}
\multiput(257.00,165.17)(0.500,-1.000){2}{\rule{0.120pt}{0.400pt}}
\put(257.0,166.0){\rule[-0.200pt]{0.400pt}{0.482pt}}
\put(258,161.67){\rule{0.241pt}{0.400pt}}
\multiput(258.00,162.17)(0.500,-1.000){2}{\rule{0.120pt}{0.400pt}}
\put(258.0,163.0){\rule[-0.200pt]{0.400pt}{0.482pt}}
\put(259,162){\usebox{\plotpoint}}
\put(259.0,160.0){\rule[-0.200pt]{0.400pt}{0.482pt}}
\put(259.0,160.0){\usebox{\plotpoint}}
\put(260.0,158.0){\rule[-0.200pt]{0.400pt}{0.482pt}}
\put(260.0,158.0){\usebox{\plotpoint}}
\put(261.0,156.0){\rule[-0.200pt]{0.400pt}{0.482pt}}
\put(261.0,156.0){\usebox{\plotpoint}}
\put(262.0,155.0){\usebox{\plotpoint}}
\end{picture}
\caption{$\z_3$ as function of $\mbox{Re}\,\gt$ with $\mbox{Im}\,\gt=0$
         for pure $\varphi^3$-theory.}
\label{figz3ofg}	 
\end{center}
\end{figure}

In \fig{figz3ofg} we present the results of a numerical calculation of $\z_3$ 
for the $\Gamma_{10}$-contour 
as function of $\gt$, as described in the Appendix. 
\begin{figure}
\begin{center}
\setlength{\unitlength}{0.240900pt}
\ifx\plotpoint\undefined\newsavebox{\plotpoint}\fi
\sbox{\plotpoint}{\rule[-0.200pt]{0.400pt}{0.400pt}}%

\caption{$\gt$ as function of $\mbox{Re}\,\gt\z_3$ 
         with $\mbox{Im}\,\gt\z_3=0$ for pure $\varphi^3$-theory.}
\label{figgofgz3}	 
\end{center}
\end{figure}
The left graph shows the real and imaginary part of $\z_3$ as function of real
and positive values of $\gt$. Notice that $\z_3(0)=1$ as demanded, and that the
imaginary part does not stay zero for real $\gt$. This is, of course, an
artifact of the definition of the path integral over a complex contour, which
is the $\Gamma_{10}$-contour for $\vhi^3$-theory in this case
(\fig{figcontours}). The right graph combines the real and imaginary part in
one curve in the complex $\z_3$-plane.

\fig{figgofgz3} shows what happens if we let $\gt$ run with real and positive
values of $g\z_3$, so that the actual coupling constant is real and positive.

%
%
\renewcommand{\gf}{g}
\subsection[Renormalization of pure $\vhi^4$-theory]%
           {Renormalization of pure $\boldsymbol{\vhi^4}$-theory}
In the case of $\vhi^4$-theory, the renormalized action is given by 
($\mu=\hbar=1$)
\begin{equation}
  S=\frac{1}{2}\z_2\vhi^2+\frac{1}{4!}\gf \z_4\vhi^4 \;\;.
\end{equation}
The \SD\ equation becomes
\begin{equation}
\z_2\phi=x-\frac{G_4}{6}(\phi^3+3\phi\phi'+\phi'')\quad,\qquad G_4=\gf\z_4\;\;,
\end{equation}
and the stepping equation \eqn{newsteppingeq}, leads to
\begin{equation}
  4G_4\frac{\partial\phi}{\partial G_4}=
  \z_2\phi''+2\z_2\phi\phi'-\phi-x\phi' \;\;,
\end{equation}
whereas \eqn{s2} assumes the form of \eqn{s2z2}.
The renormalization conditions require that

\BeginCon
\Condition\label{Con41}
$\phi(x=0)=0$;

\Condition\label{Con42}
$\phi'(0) = 1$;

\Condition\label{Con43}
$\phi'''(0)=-\gf$,
\EndCon

\noindent
and application to the \SD\ equation leads to the relation
\begin{equation}
   \z_2=1-\sixt(3-\gf)\gf\z_4 \;\;.
\label{ReEq444}   
\end{equation}
As in the case of $\vhi^3$-theory, $\phi$ can be considered to be a function of 
$x$ and $\gf$ only, and its derivative w.r.t.~$\gf$ can be written as 
\begin{equation}
\frac{\partial\phi}{\partial \gf}=
\frac{\partial\phi}{\partial \z_2} \dot{\z_2}+
\frac{\partial\phi}{\partial G_4} \dot{G_4}  \;\;.
\label{dfdg}\end{equation}
The l.h.s.~is zero in $x=0$ by \Con{Con41}, and evaluation of the r.h.s.~leads
to
\begin{equation}
   \frac{d\z_4}{d\gf} = 
   \frac{-6\z_4+(6-9\gf+3\gf^2)\z_4^2}{6\gf-\gf(6-5\gf+\gf^2)\z_4} \;\;,
\label{ReEq409}   
\end{equation}
another Abel equation of the second kind. The perturbative solution is given 
by 
\begin{equation}
   \z_4(g) = 1 + \frac{3}{2}\gf
               + \frac{3}{4}\gf^2
               + \frac{11}{8}\gf^3
               - \frac{45}{16}\gf^4
               + \frac{499}{32}\gf^5 + \cdots \;\;.
\end{equation}

We want to remark at this point that the statement by Cvitanovi\'c {\it et
al.\/}~that, in the case of $\vhi^3$-theory, the coefficients of the series
expansion of $\gt-\gt\z_3$ count connected three-point diagrams with no
self-energy or vertex insertions cannot be carried foreward to $\vhi^4$-theory:
the coefficients of the series expansion of $\gf-\gf\z_4$ do not count
connected four-point diagrams with no self-energy or (four-point) vertex
insertions. There are, for example, no such diagrams with three vertices.

To find the exact implicit solution of \eqn{ReEq409}, we apply
\Con{Con42} to the solution (\ref{SoEq440}) of $\vhi^4$-theory, resulting in
\begin{equation}
   \frac{c_1t\mbox{U}(1;t)}{\z_2}
   - \frac{c_2t\mbox{V}(1;t)}{\z_2\Gamma(\frac{3}{2})}
   = c_1\mbox{U}(0;t) 
     + \frac{c_2\mbox{V}(0;t)}{\Gamma(\half)} \quad,\qquad
t = \left(\frac{3\z_2^2}{\gf\z_4}\right)^{1/2} \;\;.
\label{phi4impl}\end{equation}
Letting 
\begin{equation}
   F_1(t) = c_1\mbox{U}(1;t)-\frac{c_2\mbox{V}(1;t)}{\Gamma(\frac{3}{2})}
   \qquad\textrm{and}\qquad
   F_0(t) = c_1\mbox{U}(0;t)+\frac{c_2\mbox{V}(0;t)}{\Gamma(\frac{1}{2})} \;,
\nn\end{equation}
the above equation becomes
\begin{equation}
   tF_1(t) = \z_2F_0(t) \;\;.
\label{ReEq410}
\nn\end{equation}
Using the properties of the parabolic functions we can easily show that
\begin{equation}
   F_1'(t) = \left(\frac{\z_2}{2}-1\right)F_0(t) \qquad,\qquad
   F_0'(t) = -\half\left(t+\frac{\z_2}{t}\right)F_0(t) \;\;,
\nn\end{equation}
so that differentiation of \eqn{ReEq410} leads to 
\begin{equation}
   \frac{dt}{d\gf}\frac{\z_2}{t}F_0(t) 
     + t\frac{dt}{d\gf}\left(\frac{\z_2}{2}-1\right)F_0(t)
   = \frac{d\z_2}{d\gf}F_0(t)
     - \frac{\z_2}{2}\frac{dt}{d\gf}\left(t+\frac{\z_2}{t}\right)F_0(t)\;\;.
\nn\end{equation}
This equation can be written as
\begin{equation}
   \frac{d}{d\gf}\frac{t}{\z_2}
   + \frac{dt}{d\gf}
     \left(\half+\frac{t^2}{\z_2^2}\!\cdot\!\frac{\gf-3}{6}\,\gf\z_4\right)
   = 0 \;\;,
\nn\end{equation}
where we used relation (\ref{ReEq444}). Finally, since
$t/\z_2=\sqrt{3/\gf\z_4}$ by definition of $t$, it is easily seen that the
above equation becomes \eqn{ReEq409}.

In \fig{figz4ofg} we show the results of the numerical calculation of
$\z_4(\gf)$, as described in the Appendix. We used the $\Gamma_{20}$-contour
for $\vhi^4$-theory (\fig{figcontours}). 
\begin{figure}
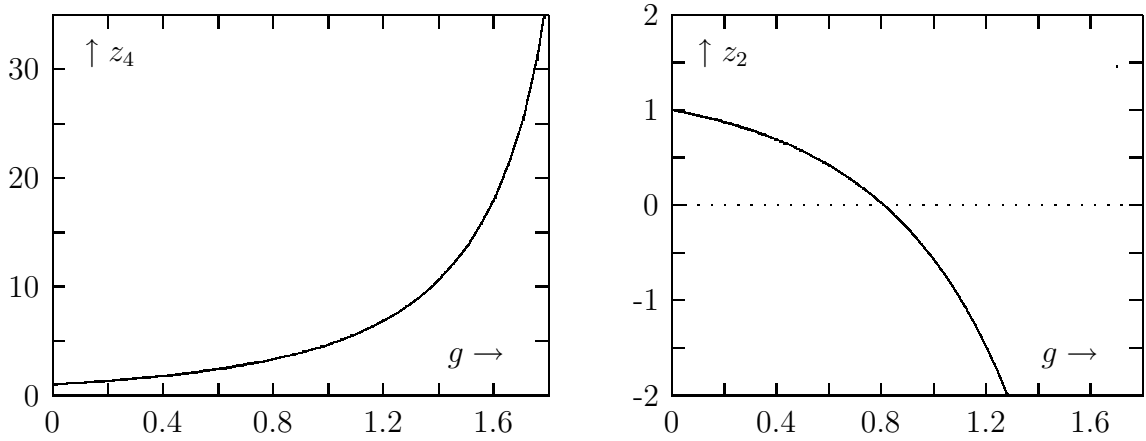

\begin{center}
\setlength{\unitlength}{0.240900pt}
\ifx\plotpoint\undefined\newsavebox{\plotpoint}\fi
\sbox{\plotpoint}{\rule[-0.200pt]{0.400pt}{0.400pt}}%

\caption{$\z_4$ as function of $\gf$ (left) and $\z_2$ as function of 
         $\gf$ (right)}
\label{figz4ofg}	 
\end{center}
\end{figure}
\begin{figure}
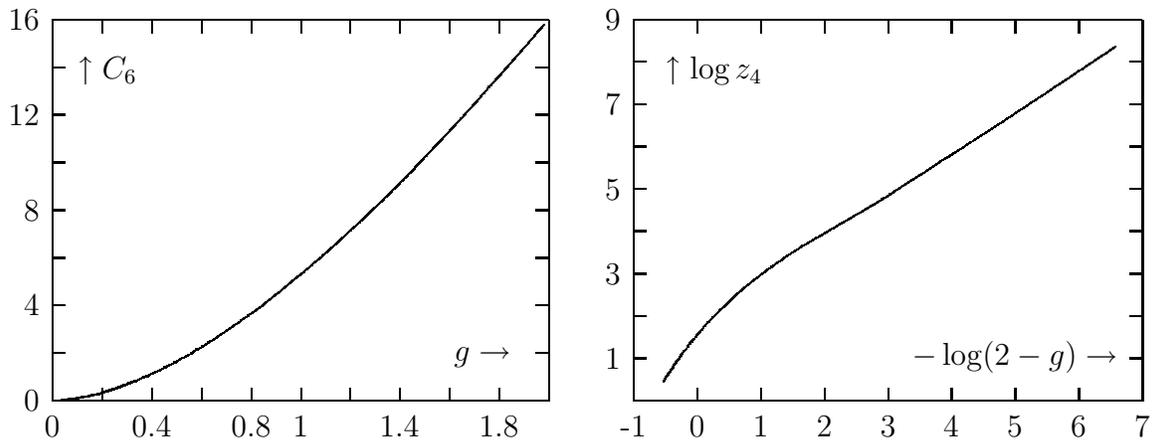

\begin{center}
\setlength{\unitlength}{0.240900pt}
\ifx\plotpoint\undefined\newsavebox{\plotpoint}\fi
\sbox{\plotpoint}{\rule[-0.200pt]{0.400pt}{0.400pt}}%
%
\caption{$C_6$ as function of $\gf$ (left) and $\log \z_4$ as function of 
$-\log(2-\gf)$ (right).}
\label{figlogz4}	 
\end{center}
\end{figure}
\begin{figure}
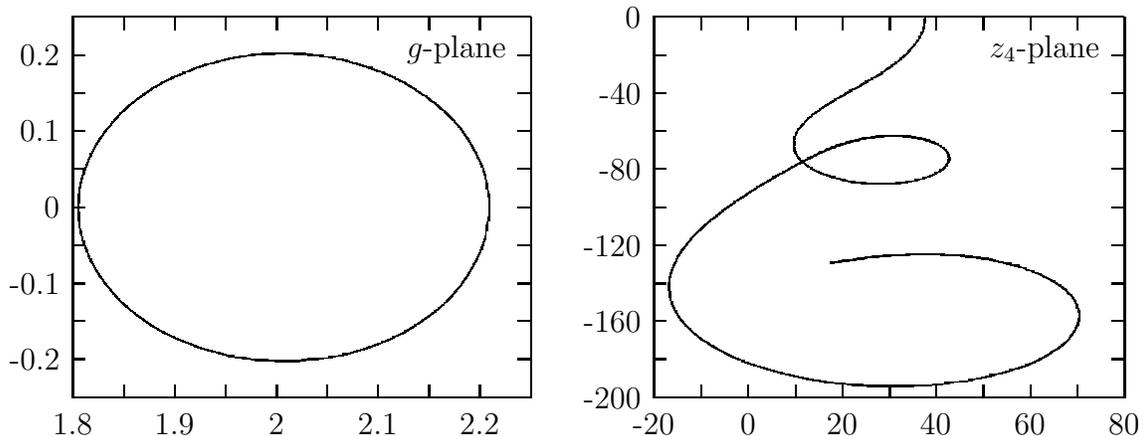

\begin{center}
\setlength{\unitlength}{0.240900pt}
\ifx\plotpoint\undefined\newsavebox{\plotpoint}\fi
%
\caption{The complex $\gf$-plane and the complex $\z_4$-plane. $\gf$ 
         goes around twice, anti-clockwise and starting on the 
	 real axis on the left side of $\gf=2$.}
\label{figgaround2}	 
\end{center}
\end{figure}
Starting at $\z_4(0)=1$, $\z_4(\gf)$ stays real and positive for real and
positive values of $\gf$, as expected. Moreover $\z_2$ exhibits a zero, whose
position $g_\star$ can be calculated analytically and is given by 
\begin{equation}
g_\star= 3-\frac{1}{4}\left(
\frac{\Gamma\left(\frac{1}{4}\right)}{\Gamma\left(\frac{3}{4}\right)}
\right)^2\sim 0.81155 \;\;.
\label{gstar}
\end{equation}
At this point the theory becomes `massless', in the sense that the bare mass
becomes zero, yet the Green's functions
 do not exhibit singular behavior.
In fact let us consider the 6-point function as an example. It can be
explicitly calculated and it reads 
\[
C_6=6\z_4^{-1}-6+9\gf+\gf^2\;\;.
\]
It is easy to see that the expansion around $\gf=0$ reproduces the known
perturbative series. Moreover, the left graph \fig{figlogz4} presents
$C_6$ as a function of $\gf$.

We also see that $\z_4$ increases with increasing $\gf$, and explodes if $\gf$
approaches $2$. The graph on the right of \fig{figlogz4} suggests that the
there is a simple pole at $\gf=2$. In fact, substitution of a Laurent series
around $\gf=2$ in \eqn{ReEq409} results in a solution with a simple pole:
\begin{equation}
  \z_4(\gf) = -\frac{6}{\gf-2} + 6 - 12(\gf-2) + 54(\gf-2)^2 - 399(\gf-2)^3 
              + 3948(\gf-2)^4 - \cdots \;.
\label{ReEq445}
\end{equation}
One can ask the question whether this series expansion corresponds to a
solution with $\z_4(0)=1$, that is, the perturbative solution. In order to get
the perturbative solution from the implicit solution (\ref{phi4impl}), in
combination with \eqn{ReEq444}, we should take the constants $c_1$ and $c_2$
such that the limit of $t\to\infty$ exists. Using the properties of the
parabolic functions and their asymptotic expansions, we find that the
perturbative solution has to satisfy
\begin{equation}
   \z_2 = t^2\left(\frac{B_{3/4}(\sfrac{1}{4}t^2)}
                         {B_{1/4}(\sfrac{1}{4}t^2)} - 1\right) \;\;,\quad
   B_\nu(\sfrac{1}{4}t^2) 
   \df \begin{cases}
          \frac{1}{\cos\nu\pi}
	  [I_{-\nu}(\frac{1}{4}t^2)+I_{\nu}(\frac{1}{4}t^2)]
	       &\textrm{if $\mbox{Re}\,t<0$} \\
	  \frac{\pi}{2\sin\nu\pi}
	  [I_{-\nu}(\frac{1}{4}t^2) - I_{\nu}(\frac{1}{4}t^2)]     
               &\textrm{if $\mbox{Re}\,t>0$,}
        \end{cases}	  
\end{equation}
together with \eqn{ReEq444}.
For $\gf$ close to, but smaller than, $\gf=2$ we see that $\z_2<0$, so that 
$\mbox{Re}\,t<0$, and it is easy to see that the solution in this case has a 
simple pole at $\gf=2$.
However, the coefficients for large powers in the series expansion seem to
behave as $(2n+2)!$, so that the series has radius of convergence equal to
zero,  and the numerical solution of a curve around $\gf=2$ in the complex
$\gf$-plane reveals that there is a branch point (\fig{figgaround2}).
In any case, for $\gf\to 2^-$ the bare coupling becomes strong and the 
bare mass squared large and negative whereas 
the connected Green's functions are
still perfectly calculable; for instance $C_6(\gf=2)=16$.
%

%
%
\renewcommand{\gt}{g_3}
\renewcommand{\gf}{g_4}
\newcommand{\Gt}{G_3}
\newcommand{\Gf}{G_4}

\subsection[Renormalization of $\vhitf$-theory]%
           {Renormalization of $\boldsymbol{\vhitf}$-theory}
The renormalization of the $\vhitf$-theory is more involved, but 
straightforward. The action is given by
\begin{equation}
  S=\frac{1}{2}\z_2\vhi^2+\frac{1}{3!}\Gt\vhi^3+\frac{1}{4!}\Gf\vhi^4 +\z_1\vhi
  \quad,\qquad \Gt=\gt\z_3 \;\;,\quad \Gf=\gf\z_4 \;\;,
\nn\end{equation}
and the \SD\ equation assumes the form
\begin{equation}
\z_2\phi=(x-\z_1)-\frac{\Gt}{2}(\phi^2+\phi')-\frac{\Gf}{6}(\phi^3
+3\phi\phi'+\phi'') \;\;.
\label{ReEq012}   
\end{equation}
The stepping equations read
\begin{align}
  \frac{\partial \phi}{\partial \Gt}&=-\frac{1}{6}\phi'''-\frac{1}{2}
  \phi\phi''-\frac{1}{2}\phi^{'2}-\frac{1}{2}\phi^2\phi' \nn\\
  \frac{\partial \phi}{\partial \Gf}&=-\frac{1}{24}\phi''''-\frac{1}{6}
  \phi\phi''-\frac{5}{12}\phi'\phi''-\frac{1}{4}\phi^2\phi''
  -\frac{1}{2}\phi\phi^{'2}-\frac{1}{6}\phi^3\phi' \nn\\
  \frac{\partial \phi}{\partial \z_2}&=-\phi\phi'-\frac{1}{2}\phi'' \nn \;\;,
\end{align}
and the renormalization conditions are now

\BeginCon
\Condition\label{Con341}
$\phi(x=0)=0$;

\Condition\label{Con342}
$\phi'(0)=1$;

\Condition\label{Con343}
$\phi''(0)=-\gt$;

\Condition\label{344}
$\phi'''(0)=3\gt^2-\gf$.
\EndCon

\noindent
Combining these conditions with the \SD\ equation one easily gets
\begin{equation}
   \z_1=\frac{1}{2}\gt\Gf-\frac{1}{2}\Gt \quad,\qquad 
   \z_2=1-\frac{1}{6}(3\gt^2-\gf+3)\Gf+\frac{1}{2}\gt\Gt \;\;,
\end{equation}
so that $\phi$ becomes a function of $\gt$, $\gf$ and $x$ only, leading to the 
the four equations:
\begin{equation}
\left.\frac{\partial \phi}{\partial g_i}\right|_{x=0}\equiv
-\phi'(0)\frac{\partial \z_1}{\partial g_i}+
\left.\frac{\partial \phi}{\partial \z_2}\right|_{x=0}
\frac{\partial \z_2}{\partial g_i}+
\left.\frac{\partial \phi}{\partial \Gt}\right|_{x=0}
\frac{\partial \Gt}{\partial g_i}+
\left.\frac{\partial \phi}{\partial \Gf}\right|_{x=0}
\frac{\partial \Gf}{\partial g_i}=0
\nn\end{equation}
and
\begin{equation}
\left.\frac{\partial \phi'}{\partial g_i}\right|_{x=0}\equiv
-\phi''(0)\frac{\partial \z_1}{\partial g_i}+
\left.\frac{\partial \phi'}{\partial \z_2}\right|_{x=0}
\frac{\partial \z_2}{\partial g_i}+
\left.\frac{\partial \phi'}{\partial \Gt}\right|_{x=0}
\frac{\partial \Gt}{\partial g_i}+
\left.\frac{\partial \phi'}{\partial \Gf}\right|_{x=0}
\frac{\partial \Gf}{\partial g_i}=0
\nn\end{equation} 
with $i=3,4$. 
The coefficients at $x=0$ can be inferred form the stepping equations.
This way we have a system of four equations involving the partial
derivatives of the functions $\Gt(\gt,\gf)$ and $\Gf(\gt,\gf)$ with respect
to $\gt$ and $\gf$. Notice that the equations are linear with respect to
the four partial derivatives but higly non-linear with respect to the
functions $\Gt(\gt,\gf)$ and $\Gf(\gt,\gf)$.
They can be solved perturbatively with the result
\begin{eqnarray*}
\Gt&=&
 \gt - \gt^3\,\hbar + 
   \frac{3}{2}\, \gt\, \gf\,\hbar +
   4\,\gt^5\,\hbar^2 
- 6\,\gt^3\, \gf\,\hbar^2 + 
   \frac{3}{4}\, \gt\,\gf^2\,\hbar^2 - 
   4\,\gt^7\,\hbar^3 - 
   \frac{3}{2}\,\gt^5\, \gf\,\hbar^3 
\\
&+&
   \frac{19}{4} \,\gt^3\,\gf^2\,\hbar^3+ 
   \frac{11}{8}\, \gt\,\gf^3\,\hbar^3 + 
   7\,\gt^9\,\hbar^4 - 
   \frac{93}{2} \,\gt^7\, \gf\,\hbar^4+ 
   81\,\gt^5\,\gf^2\,\hbar^4 - 
   \frac{100}{3}\,\gt^3\,\gf^3\,\hbar^4 
\\
&-& 
   \frac{45}{16}\, \gt\,\gf^4\,\hbar^4 + 
   47\,\gt^{11}\,\hbar^5 - 
   \frac{807}{2}\,\gt^9\, \gf\,\hbar^5 + 
   927\,\gt^7\,\gf^2\,\hbar^5 - 
   \frac{2787}{4}\,\gt^5\,\gf^3\,\hbar^5 
\\
&+& 
   \frac{1785}{16}\,\gt^3\,\gf^4\,\hbar^5 + 
   \frac{499}{32}\, \gt\,\gf^5\,\hbar^5
\\
\Gf
&=&
  \gf + 3\,\gt^4\,\hbar - 
   6\,\gt^2\, \gf\,\hbar + 
   \frac{3}{2} \,\gf^2\,\hbar
 - 6\,\gt^6\,\hbar^2 + 
   5\,\gt^4\, \gf\,\hbar^2 + 
   \frac{3}{2} \,\gt^2\,\gf^2\,\hbar^2+ 
   \frac{3}{4} \,\gf^3\,\hbar^2
\\
&+& 9\,\gt^8\,\hbar^3 - 
   43\,\gt^6\, \gf\,\hbar^3 + 
   \frac{151}{2}\,\gt^4\,\gf^2\,\hbar^3 - 
   39\,\gt^2\,\gf^3\,\hbar^3 + 
   \frac{11}{8} \,\gf^4\,\hbar^3+ 
33\,\gt^{10}\,\hbar^4 
\\
&-& 
   324\,\gt^8\, \gf\,\hbar^4 + 
   834\,\gt^6\,\gf^2\,\hbar^4 - 
   \frac{1485}{2}\,\gt^4\,\gf^3\,\hbar^4 + 
   \frac{1585}{8}\,\gt^2\,\gf^4\,\hbar^4 - 
   \frac{45}{16} \,\gf^5\,\hbar^4
\\
&+& 
   \frac{1029}{2}\,\gt^{12}\,\hbar^5 - 
   4610\,\gt^{10}\, \gf\,\hbar^5 + 
   \frac{27525}{2}\,\gt^8\,\gf^2\,\hbar^5 - 
   17020\,\gt^6\,\gf^3\,\hbar^5 + 
   \frac{68595}{8}\,\gt^4\,\gf^4\,\hbar^5 
\\
&-& 
   \frac{10705}{8} \,\gt^2\,\gf^5\,\hbar^5+ 
   \frac{499}{32}\,\gf^6\,\hbar^5
\end{eqnarray*}
where the $\hbar$ dependence has been restored for convenience.

In the limit $\gt \to 0$, also $\Gt \to 0$, and the equations reduce to 
\begin{alignat}{2}
  \frac{\partial \Gf}{\partial \gt}&=0 &\quad,\qquad
  \frac{\partial \Gf}{\partial \gf}&=
  \frac{2(2-\gf)\Gf^2}{6 \gf-(6-5 \gf+\gf^2)\Gf} \;\;,\nn 
  \\
  \frac{\partial \Gt}{\partial \gt}&=\frac{\Gf}{\gf} &\quad,\qquad
  \frac{\partial \Gt}{\partial \gf}&=0 \;\;.\nn
\end{alignat}
Note that $\Gf(0,\gf)$ can be identified as $\gf\z_4(\gf)$ where
$\z_4$ is the same function is as in pure $\vhi^4$-theory.
Another interesting result is that the term linear in $\gt$ in the expansion of 
$\Gt$ is given by
\[ 
\Gt(\gt,\gf)=\gt \z_4(\gf)+{\cal O}(\gt^2)\;\;.
\]
%
%
\newcommand{\psib}{\bar{\psi}}
\newcommand{\phib}{\bar{\phi}}
\newcommand{\frth}{\frac{1}{4}}
\newcommand{\uu}{u}
\newcommand{\ub}{\bar{\uu}}
\renewcommand{\gg}{g}
\newcommand{\ddg}[1]{\frac{d #1}{d\gg}}
\newcommand{\srac}[2]{{\textstyle\frac{#1}{#2}}}
\newcommand{\UU}{\mbox{U}}

\subsection{Renormalization of the charged scalar field}
In the case of the charged scalar field we consider the integral
representation 
\begin{equation}
   R(\xx,\xb) = \sqrt{\frac{\mm}{\hbar}}\int d\vhi d\vhib\,\exp
   \left\{-\frac{1}{\hbar}\left[\mm\z_2\vhi\vhib + \frth\la\z_4(\vhi\vhib)^2 
                                - \xx\vhib - \xb\vhi\right]\right\} \;\;,
\label{ReEq017}				
\end{equation}
which satisfies the \SD\ equation
\begin{equation}
   \zeta R'''(\zeta) + 2R''(\zeta) + \frac{2}{\gg\z_4}[\z_2R'(\zeta)-R(\zeta)]
   = 0 \quad,\qquad
   \zeta = \frac{\xx\xb}{\mm\hbar} \;\;,\quad
   \gg = \frac{\la\hbar}{\mm^2} \;\;.
\label{ReEq018}   
\end{equation}
In the dimensionless variables
\begin{equation}
   \uu = \frac{\xx}{\sqrt{\mm\hbar}} \;\;,\quad
   \ub = \frac{\xb}{\sqrt{\mm\hbar}} \;\;,\quad
   \zeta = \uu\ub \;\;,\quad
   \psi = \sqrt{\frac{\mm}{\hbar}}\,\vhi \;\;,\quad
   \psib = \sqrt{\frac{\mm}{\hbar}}\,\vhib \;\;,
\nn\end{equation}
\eqn{ReEq017} becomes 
\begin{equation}
   R(\uu,\ub) = \int d\psi d\psib\,\exp
   \left(-\z_2\psi\psib - \frth\gg\z_4(\psi\psib)^2 + \uu\psib 
                                                    + \ub\psi\right) \;\;,
\nn\end{equation}
implying
\begin{equation}
   \frac{\pa R}{\pa\gg} = - \frth\ddg{(\gg\z_4)}
\frac{\pa^4\!R}{\pa\uu^2\pa\ub^2}
                          - \ddg{\z_2}\frac{\pa^2\!R}{\pa\uu\pa\ub}
\nn\end{equation}
or, in terms of the $\zeta$-variable
\begin{equation}
   \frac{\pa R}{\pa\gg} = 
   -\frth\ddg{(\gg\z_4)}(\zeta^2R'''' + 4\zeta R''' + 2R'') 
   - \ddg{\z_2}(\zeta R'' + R') \;\;.
\label{ReEq020}   
\end{equation}
%
%
Now, the generating function of the connected Green's functions is given by 
\begin{equation}
   \phi(\xx,\xb) = \hbar\frac{\pa}{\pa\xb}\ln R(\zeta) 
                 = \frac{\xx}{\mm}\frac{R'(\zeta)}{R(\zeta)}  \;\;,
\nn\end{equation}
and the renormalization conditions are

\BeginCon
\Condition\label{ConC1}
${\displaystyle\frac{\pa\phi}{\pa\xx}(\xx=\xb=0)=\frac{1}{\mm}}$ , implying 
$R'(0)=R(0)$ ;

\Condition\label{ConC2}
${\displaystyle\frac{\pa^3\phi}{\pa\xb\pa\xx^2}(\xx=\xb=0)
=-\frac{\la}{\mm^4}}$ , implying 
${\displaystyle R''(0) = \left(1-\frac{\gg}{2}\right)R(0)}$ .
\EndCon

\noindent
By combining equations 
\eqn{ReEq018} and \eqn{ReEq020}  and the renormalization conditions
we get
\begin{equation}
  \z_2 = 1 - \left(1-\frac{\gg}{2}\right)\gg\z_4 
\nn\end{equation}
and
\begin{equation}
   \ddg{\z_4} 
   = \frac{2\z_4 - (3\gg^2-5\gg+2)\z_4^2}
          {-2\gg + \gg(\gg^2-3\gg+2)\z_4} \;\;,
\label{ReEq025}	  
\end{equation}
with perturbative expansion
\begin{equation}
   \z_4(\gg) = 1 + \frac{5}{2}\gg + \frac{9}{4}\gg^2 
                 + \frac{49}{8}\gg^3
                 - \frac{271}{16}\gg^4 + \frac{5025}{32}\gg^5 + \cdots \;\;.
\end{equation}
To get an exact implicit solution of \eqn{ReEq025}, we go back to \eqn{ReEq018} and
change variables to get
\begin{equation}
   \eta R'''(\eta) + 2R''(\eta) + \al^2[R'(\eta)-R(\eta)] = 0 \quad,\qquad
   \eta = \frac{\zeta}{\z_2} \;\;,\quad
   \al = \sqrt{\frac{2}{\gg\z_4}}\,\z_2 \;\;.
\label{ReEq026}   
\end{equation}
This equation has exactly the form of (\ref{SoEq064}), and the perturbative
solution is given by 
\begin{equation}
   R(\eta) = \sum_{n=0}^\infty\frac{(\eta\al)^n}{n!}\,
                              \UU(n+\srac{1}{2};\al) \;\;.
\nn\end{equation}
\Con{ConC1} implies the implicit exact solution of the form
\begin{equation}
   \al\UU(\srac{3}{2};\al) = \z_2\UU(\srac{1}{2};\al) \;\;,
\label{ReEq027}   
\end{equation}
where, of course, $\z_2 = 1 - (1-\gg/2)\gg\z_4$. 

To show that \eqn{ReEq027} is 
indeed an implicit solution of \eqn{ReEq025}, 
differentiate \eqn{ReEq027} with respect to $\gg$:
\begin{equation}
   [\UU(\srac{3}{2};\al)+\al\UU'(\srac{3}{2};\al)]\ddg{\al}
   = \ddg{\z_2}\UU(\srac{1}{2};\al) + \z_2\UU'(\srac{1}{2};\al)\ddg{\al} \;\;,
\nn\end{equation}
and using parabolic cylinder functions properties
together with \eqn{ReEq027} to get
\begin{equation}
   \ddg{\z_2} = \left(\frac{\z_2}{\al} + \al\z_2 + \frac{\z_2^2}{\al} 
                                               - \al\right)\ddg{\al} \;\;,
\end{equation}
with $\al = \sqrt{\frac{2}{\gg\z_4}}\,\z_2$. It is a straight forward 
calculation to show that this is indeed \eqn{ReEq025}.

In the following we present a derivation of the  initial condition 
for $\z_4(\gg=0)$. 
Using the path integral expression of \eqn{ReEq017}, we find the \SD\ equation
\begin{equation}
   \mm\z_2\hbar\pb R + \frac{\la\z_4}{2}\hbar^3\pa\pb^2 R - xR = 0 \;\;.
\nn\end{equation}
The generating function $\phi=\hbar\pb\ln R$ of the connected diagrams
satisfies
\begin{equation}
   \phi(0,0) = 0 \;\;,\quad
   (\pb\phi)(0,0)=0 \;\;,\quad
   \phib(0,0) = 0 \;\;,\quad
   (\pa\phib)(0,0) = 0 \;\;.
\nn\end{equation}
as well as the \SD\ equation
\begin{equation}
   \phi = \frac{\xx}{\mm\z_2} 
          - \frac{\la\z_4}{2\mm\z_2}(\phib\phi^2 + 2\hbar\phi\pa\phi 
	                            + \phib\pb\phi + \hbar^2\pb\pa\phi)\;\;,
\nn\end{equation}
with the renormalization conditions \ref{ConC1} and \ref{ConC2}. These should 
hold for any value of $\hbar$, and for $\hbar=0$, the \SD\ equation becomes
\begin{equation}
   \phi_0 = \frac{\xx}{\mm\z_2(0)} 
            - \frac{\la\z_4(0)}{2\mm\z_2(0)}\phib_0\phi_0^2 \;\;,
\nn\end{equation}
from which we derive for the perturbative solution that
\begin{alignat}{2}
   (\pa\phi_0)(0,0) &= \frac{1}{\mm\z_2(0)} \quad&\Rightarrow\quad
   \z_2(0) &= 1 \;\;,\nn \\
   (\pb\pa^2\phi_0)(0,0) &= -\frac{\la\z_4(0)}{\mm^4} \quad&\Rightarrow\quad
   \z_4(0) &= 1 \;\;.
\nn\end{alignat}
Notice that the value $\hbar=0$ is directly related to $g=0$ since 
$g$ is proportional to $\hbar$.

\section{Summary}

In this paper we studied several aspects of zero-dimensional field theories.
In the first place we derived a set of diagrammatic equations, including the
well known Schwinger-Dyson equations as well as a set of `stepping' equations
generalizing some previous results. Then we showed how to solve these equations
exactly in terms of known functions and we established integral representations
of these solutions, best known as the `path integral' representation. Explicit
results were obtained for $\vhi^3$, $\vhi^4$, $\vhi^3+\vhi^4$ and the charged
scalar field theories. Subsequently, we studied the `renormalization' of such
theories in zero dimensions, which is equivalent to counting diagrams with
restrictions imposed on the type of diagrams considered, for instance diagrams
without any tadpoles, self-energy insertions or vertex insertions. We were able
to get explicit results for the dependence of the bare quantities such as the
mass, the coupling, and the tadpole counter terms, on the renormalized
(physical) coupling constant. Examples of interesting observations are the
facts that in $\vhi^4$ theory, the bare mass exhibits a zero at a finite value
of the renormalized coupling constant $g=g_\star$ (\eqn{gstar}), whereas at
$g\to 2-\epsilon$ the bare coupling becomes strong and the mass squared becomes
large and negative. Yet in both cases the `physical' connected Green's
functions remain finite and calculable.

\newpage
\section*{Appendix}

Consider general $\vhi^p$-theory, and suppose that all but one renormalization
conditions have been implemented through functions $z_k(g,z_p)$, $k=1,2,..,p-1$
of two variables $z_p$ and $g$, like in \eqn{ReEq333} and \eqn{ReEq444}. 
This means that we are considering a theory with an action
\begin{equation}
   S(g,z_p;\vhi) = \frac{gz_p}{p!}\,\vhi^p 
           + \sum_{k=1}^{p-1}\frac{z_k(g,z_p)}{k!}\,\vhi^k 
   \;\;.
\nn\end{equation}
Let $\Gamma$ be a contour in the complex $\vhi$-plane, such that 
$\mbox{Re}\,\vhi^n\to\infty$ at the endpoints, and define
\begin{equation}
   Z_n(g,z_p) 
   \df \int_{\Gamma}d\vhi\,\vhi^n\,\exp\{-S(g,z_p;\vhi)\} \;\;.
\nn\end{equation}
Such an integral is not defined for all complex values of $gz_p$.  Let
$gz_p=|gz_p|e^{ip\eta}$ and denote by $e^{-i\eta}\Gamma$ the contour that is
obtained from $\Gamma$ by clockwise rotation over $\eta$. For complex values
of $gz_p$, we define
\begin{align}
   Z_n(g,z_p) 
   \df &\;\int_{e^{-i\eta}\Gamma}d\vhi\,\vhi^n\,\exp\{-S(g,z_p;\vhi)\} \notag\\
   = &\;e^{-i(n+1)\eta}\int_{\Gamma}d\vhi\,\vhi^n\,
       \exp\left( - \frac{|gz_p|}{p!}\,\vhi^p 
                  - \sum_{k=1}^{p-1}\frac{z_k}{k!}\,e^{-ik\eta}\vhi^k\right)
		  \;\;.
\nn\end{align}
Integrals of this type can easily be calculated to high precision by numerical 
integration. One just has to choose $\Gamma$ such that it goes through one or 
more saddle points, so that the integrand oscillates as little as possible.

To formulate the renormalization problem further, let us denote the connected
moments by $C_n$, so
\begin{equation}
   C_1=\frac{Z_1}{Z_0} \;\;,\quad
   C_2=\frac{Z_2}{Z_0} - C_1^2 \;\;,\quad
   C_3=\frac{Z_3}{Z_0} - 3C_2C_1 - C_1^3 \;\;,\ldots
\nn\end{equation}
and so on. The problem is to solve $z_p$ as function of $g$ from the implicit
function equation
\begin{equation}
   C_p(g,z_p) = - g \;\;,
\nn\end{equation}
which represents the final renormalization condition. This equation can be
solved numerically. Given a value of $g$, we have to find the zero of the
function
\begin{equation}
   F(z_p) \df C_p(g,z_p) + g \;\;,
\nn\end{equation}
which can be found using Newton-Raphson iteration
\begin{equation}
   z_p \leftarrow z_p - \frac{F(z_p)}{F'(z_p)} \;\;.
\nn\end{equation}
By making small steps in the value of $g$, the solution $z_p(g)$ on a curve in
the complex $g$-plane can be determined. At the start of each iteration, the
question arises of which initial value of $z_p$ to choose, and the obvious
answer is to choose the final value of the previous iteration, which should lie
close the the new final value if the steps in $g$ are not to large. 

As a check one can look whether the results obtained with this method satisfy
(numerically) the available differential equations for $z_p(g)$ (\eqn{ReEq005}
and \eqn{ReEq409}).

\end{document}